\documentclass[10pt,journal]{IEEEtran}
\IEEEoverridecommandlockouts % show footnote comments
\usepackage{blkarray}                                      % to support matrix
\usepackage{algpseudocode}                                 % new algorithm package
\usepackage{algorithm}
\usepackage{graphicx}                                      % include pdf figures
\usepackage{amsmath}
\usepackage{amssymb}
\usepackage{amsfonts}
\usepackage{amsthm}
\usepackage[mathcal]{eucal}
\usepackage{mathrsfs}
\usepackage{booktabs}
\usepackage{enumerate}
\usepackage{multirow}
\usepackage{color}
\usepackage{cite}                                          % more citations in one bracket
\usepackage{comment}                                       % use comment
\usepackage{soul}                                          % use highlight command \hl{}
\soulregister\cite7
\soulregister\ref7
\soulregister\pageref7
\usepackage{etoolbox}                                      % commands \newtoggle, \toggletrue, \iftoggle
\usepackage{url}
\usepackage{nth}                                           % nth command
\usepackage{bm}                                            % bm command
\usepackage{courier}
\usepackage{balance}
\usepackage{threeparttable}
\usepackage{xcolor,colortbl}
\usepackage{footnote}
\usepackage{listings}
\usepackage{setspace}                                      % setstretch{}
\usepackage{verbatim}
\usepackage[bookmarks=false]{hyperref}
\hypersetup{
    colorlinks = true,
    citecolor  = blue,
    linkcolor  = blue,
    urlcolor   = blue,
}
\usepackage{tikz}
\usetikzlibrary{patterns,snakes}
\usetikzlibrary{positioning,calc,fit,decorations.pathmorphing,shapes.geometric,shapes.gates.logic.US, calc}
\usetikzlibrary{arrows,arrows.meta,decorations.markings,shapes,shapes.arrows,shadows}
\usetikzlibrary{perspective}
\usetikzlibrary{decorations,decorations.pathreplacing,decorations.text}
\usetikzlibrary{backgrounds}
\usepackage{filecontents}                                  % support to pgfplots
\usepackage{pgfplots}
\usepackage{pgfplotstable}
\usepackage{scalefnt}
\pgfplotsset{compat=newest}
\usepackage{caption}
\usepackage{subcaption} % required for subref
\usepackage{cleveref}
\Crefformat{figure}{Fig.~#2#1#3}                           % "Fig.", instead of "Figure"
\Crefname{subfigure}{Fig.}{Figs.}
\Crefname{figure}{Fig.}{Figs.}
\Crefformat{table}{TABLE~#2#1#3}                           % "TABLE", instead of "Table"
\usepackage[figuresright]{rotating}
\definecolor{CUHKorange}{RGB}{244,106,18} %F47012
\definecolor{CUHKblue}{RGB}{0,111,190}    %006FBE
\definecolor{CUHKgreen}{RGB}{0,127,128}   %007F80
\definecolor{CUHKred}{RGB}{228,46,36}     %E42E24
\definecolor{CUHKyellow}{RGB}{198,148,34} %C69422
\definecolor{CUHKdark}{RGB}{114,44,114}   %722C72
\definecolor{CUHKmiddle}{RGB}{144,44,144} %902C90
\definecolor{CUHKlight}{RGB}{167,44,167} 
\definecolor{CUHKpurple}{RGB}{117,15,109}
\definecolor{CUHKgold}{RGB}{221,163,0}
\definecolor{CUHKribbon}{RGB}{244,223,176}
\definecolor{CUHKblack}{RGB}{34,24,21}
\definecolor{slategray}{RGB}{112,128,144}
\definecolor{darkgray}{RGB}{47,79,79}
\definecolor{rosybrown}{RGB}{188,143,143}
\definecolor{deepblue}{RGB}{64,115,158}

\renewcommand{\bf}[1]{\textbf{#1}}
\renewcommand{\tt}[1]{\texttt{#1}}
\DeclareMathOperator*{\fdiff}{\Delta}

\newtheorem{definition}{Definition}
\newtheorem{theorem}{Theorem}
\newtheorem{corollary}{Corollary}

\newtheorem{assumption}{Assumption}

\crefname{theorem}{Theorem}{Theorems}
\crefname{lemma}{Lemma}{Lemmas}
\crefname{property}{Property}{Properties}
\crefname{corollary}{Corollary}{Corollaries}
\crefname{assumption}{Assumption}{Assumptions}

\usepackage{tcolorbox}
\tcbuselibrary{skins,breakable}
    {\endtcolorbox}

\usepackage[top=0.75in,bottom=0.80in,left=0.58in,right=0.58in]{geometry}
\crefname{mytheorem}{Theorem}{Theorems}
\crefname{mylemma}{Lemma}{Lemmas}
\crefname{myclaim}{Claim}{Claims}
\crefname{myproperty}{Property}{Properties}
\crefname{mycorollary}{Corollary}{Corollaries}

\algrenewcommand\textproc{\texttt}

\makeatletter
\let\OldStatex\Statex
\renewcommand{\Statex}[1][3]{%
  \setlength\@tempdima{\algorithmicindent}%
  \OldStatex\hskip\dimexpr#1\@tempdima\relax
}
\makeatother

\RequirePackage[normalem]{ulem} %DIF PREAMBLE
\RequirePackage{color}\definecolor{RED}{rgb}{1,0,0}\definecolor{BLUE}{rgb}{0,0,1} %DIF PREAMBLE
\graphicspath{{./figs/}{../}}
\usepackage{fontawesome}
\usepackage{calc}
\usepackage{lipsum}
\usepackage{siunitx}
\usepackage{tikz-3dplot}

\newcommand{\ubold}{\fontseries{b}\selectfont} 
\robustify\ubold

\newcommand{\power}[2][10]{#1\textsuperscript{#2}}
\allowdisplaybreaks

\begin{document}

\twocolumn
\title{
    Analytical Die-to-Die 3D Placement with Bistratal Wirelength Model and GPU Acceleration
}

\author{
  Peiyu Liao\textsuperscript{$\dag$}, \quad
  Yuxuan Zhao\textsuperscript{$\dag$}, \quad
  Dawei Guo, \quad
  Yibo Lin, \quad 
  Bei Yu
  \thanks{\textsuperscript{$\dag$}These authors contributed equally.}
  \thanks{This work is supported in part by The Research Grants Council of Hong Kong SAR (Project No.~CUHK14208021).}
  \thanks{This work has been submitted to the IEEE for possible publication. Copyright may be transferred without notice, after which this version may no longer be accessible.}
  \thanks{P.~Liao, Y.~Zhao, and B.~Yu are with the Department of Computer Science and Engineering, The Chinese University of Hong Kong, NT, Hong Kong SAR.}
  \thanks{D.~Guo and Y.~Lin are with the School of Integrated Circuits, Peking University, China}
}

\maketitle

\begin{abstract}
In this paper, we present a new analytical 3D placement framework with a bistratal wirelength model for F2F-bonded 3D ICs with heterogeneous technology nodes based on the electrostatic-based density model. 
The proposed framework, enabled GPU-acceleration, is capable of efficiently determining node partitioning and locations simultaneously, leveraging the dedicated 3D wirelength model and density model.
The experimental results on ICCAD 2022 contest benchmarks demonstrate that our proposed 3D placement framework can achieve up to 6.1\% wirelength improvement and 4.1\% on average compared to the first-place winner with much fewer vertical interconnections and up to 9.8$\times$ runtime speedup. Notably, the proposed framework also outperforms the state-of-the-art 3D analytical placer by up to 3.3\% wirelength improvement and 2.1\% on average with up to 8.8$\times$ acceleration on large cases using GPUs.
\end{abstract}

\section{Introduction}
\label{sec:intro}
\IEEEPARstart{W}{ith} technology scaling nearing its physical limits, the 3D integrated circuit (3D-IC) has emerged as a promising solution for extending Moore's Law.
Vertically stacking multiple dies enables 3D-IC to achieve higher transistor density and replace long 2D interconnects with shorter inter-die connections, leading to improved circuit performance.
Leveraging advanced packaging technology, chiplets with heterogeneous technology nodes can be integrated to achieve leading cost-effective performance.
Prominent examples of such technology adoption are Intel's Meteor Lake~\cite{gomes2022meteor} and AMD's Zen 4~\cite{munger2023zen}, which have resulted in significant performance gains and cost savings.

Conventionally, 3D-ICs are fabricated using through-silicon vias (TSVs) with large pitches and parasitics, which may limit the total number of global interconnects to avoid performance degradation~\cite{dong2010fabrication}.
As an alternative approach, monolithic 3D (M3D) integration has been proposed, where tiers are fabricated sequentially and connected using monolithic inter-tier vias (MIVs)~\cite{batude20123,samal2016monolithic,panth2017shrunk,ku2018compact}.
In contrast to TSVs with microscale pitches, MIVs exhibit nanoscale dimensions~\cite{samal2016monolithic}, allowing for higher integration density with significantly reduced space requirements. 
Nevertheless, it is still necessary to allocate certain white space on placement regions to accommodate MIVs.
Face-to-face (F2F) bonding is another approach that bonds ICs using face sides for both dies~\cite{morrow2006three,jung2014enhancing,panth2014design,song2015coupling}.
F2F-bonded 3D ICs do not require additional silicon area for 3D connections~\cite{jung2014enhancing}, eliminating the need to reserve white space for vias and allowing much higher integration density. 
The silicon-space overhead-free property of F2F-bonded 3D ICs provides significant advantages in numerous applications~\cite{ku2018compact}.

The emergence of 3D-IC presents challenges to traditional 2D electronic design automation methods in producing high-quality 3D circuit layouts, and the heterogeneous technology nodes further complicates the problem.
Placement plays a dominant role on the overall quality of physical design, and innovations of 3D-IC placement are required to fully benefit from the 3D integration technologies.
Within the context of 3D placement, 3D-IC placers are responsible for solving the optimal 3D node locations to optimize specific objectives. Such a very large-scale combinatorial optimization problem can be solved in either discrete or analytical algorithms. 
An analytical 3D placement algorithm is characterized by employing ``true-3D'' flows that handle tier partitioning continuously and devise 3D solutions directly.
%A 3D placement algorithm is considered to be analytical if and only if it applies ``true-3D'' flows that handle tier partitioning continuously and devise 3D solutions directly.

Despite the various research achievements mentioned above, existing discrete and analytical 3D flows are hardly applicable to F2F-bonded 3D ICs with heterogeneous technology nodes.
The discrete solutions typically fail to utilize the advantages of 3D ICs sufficiently as most of them rely on the FM-mincut tier partitioning~\cite{fiduccia1988linear}.
However, the total cutsize is not the primary placement objective in F2F-bonded 3D ICs due to the silicon-space overhead-free property~\cite{jung2014enhancing}, resulting in sub-optimal partitioning for discrete solutions.
Conventional analytical 3D placement algorithms adopt continuous optimization but they do not support heterogeneous technology nodes during global placement. 
Additionally, previous wirelength-driven analytical placement algorithms use inaccurate wirelength models~\cite{lu2016eplace} for numerical optimization, which is inconsistent with F2F-bonded scenarios.
Some recent work~\cite{chen2023analytical} on wirelength models supports heterogeneous technology nodes in analytical placement. 
However, it still pays no attention to the wirelength reduction introduced by inter-die connections, remaining unsolved inaccurate estimation in 3D analytical placement.

In this paper, we propose a new analytical 3D placement framework for F2F-bonded 3D ICs with heterogeneous technology nodes utilizing a novel and precise bistratal wirelength model.
Based on the proposed placement framework, we efficiently determine the node locations along with partitioning in a single run.
The main contributions are summarized as follows.
\begin{itemize}
\item We design a \emph{bistratal wirelength} model, including computation strategies of the wirelength objective and gradients, that significantly outperforms the widely-used models for F2F-bonded 3D ICs.
\item We propose an ultra-fast analytical 3D placement framework that leverages the \emph{bistratal wirelength} model and eDensity-3D~\cite{lu2016eplace} with GPU acceleration, considering heterogeneous technology nodes.
\item Experimental results show that our results achieved the best results on the ICCAD 2022 Contest Benchmarks~\cite{hu20222022} with up to 6.1\% wirelength improvement and 4.1\% on average, compared to the first-place winner. Remarkably, we also outperform the state-of-the-art (SOTA) analytical 3D placer~\cite{chen2023analytical} for heterogeneous F2F-bonded 3D ICs by up to 3.3\% wirelength improvement and 2.1\% on average. The usage of vertical interconnects are also significantly reduced.
\end{itemize}

The rest of this paper is structured as follows.
\Cref{sec:prelim} provides some preliminaries, including previous works and foundations of analytical placement.
\Cref{sec:problem-formulation} discusses the problem statement and problem formulation.
\Cref{sec:overall-placement-flow} presents the overall flow of the proposed placement framework for heterogeneous F2F-bonded 3D ICs.
Then,~\Cref{sec:wirelength-model} depicts the theoretical details of the bistratal wirelength model.
~\Cref{sec:results} presents experimental results and some related analysis on the adopted benchmarks, followed by the conclusion in~\Cref{sec:conclusion}.

\section{Preliminaries}
\label{sec:prelim}

\subsection{Related Works}
\label{subsec:related-works}
Conventional discrete solutions handle multiple tiers discretely. {T3Place}~\cite{cong2007thermal} transforms 2D placement solutions into 3D with several folding techniques and local refinement.
Early TSV-based research on partition-based approaches~\cite{deng2001interconnect,das2003design,goplen2007placement,kim2009study} first partitions the netlist to minimize specific targets, \emph{e.g.}, vertical connections, followed by a simultaneous 2D placement on all tiers.
The ``pseudo-3D'' flows utilize optimization techniques of existing 2D engines to work with projected 3D designs.
{Cascade2D}~\cite{chang2016cascade2d} implements an M3D design using 2D commercial tools with a design-aware partitioning before placement. 
Recent partitioning-based approaches~\cite{chang2016cascade2d,panth2017shrunk,ku2018compact,park2020pseudo} suggest that partitioning first may not sufficiently leverage physical information and thus perform partitioning-last strategies after 2D pre-placement.
{Shrunk-2D}~\cite{panth2014design,panth2017shrunk} is a prominent example that performs partitioning according to a 2D pre-placement. {Shrunk-2D} requires geometry shrinking of standard cells and related interconnects by 50\% during its 2D pre-placement for F2F-bounded 3D ICs~\cite{panth2014design} or M3D~\cite{panth2017shrunk}.
{Compact-2D}~\cite{ku2018compact} adopts placement contraction without geometry shrinking to obtain the 2D pre-placement, followed by a bin-based FM-mincut tier partitioning~\cite{fiduccia1988linear}.
{Pin-3D}~\cite{pentapati2020pin} proposes pin projection to incorporate inter-die physical information by projecting pins to other dies with fixed locations and transparent geometries, which is first applicable to heterogeneous monolithic 3D ICs.
{Snap-3D}~\cite{vanna2021snap} for F2F bonded 3D ICs shrinks the height of standard cell layouts by one half and labels footprint rows top vs.~bottom to indicate partitioning.
However, the bin-based min-cut partitioning algorithm lacks an understanding of the impact of partitioning on placement quality.
{TP-GNN}~\cite{lu2020tp}, an unsupervised graph-learning-based tier partitioning framework, is proposed to address this drawback for M3D ICs using graph neural networks (GNNs).
Considering that discrete algorithms are particularly sensitive to partitioning~\cite{murali2022art} and can potentially lead to performance degradation, analytical 3D placement is considered to be more promising to produce solutions with higher quality.

Analytical 3D solutions relax discrete tier partitioning and solve continuous 3D optimization problems.
Typical analytical approaches include quadratic programming~\cite{kaya20033,hentschke2006quadratic}, nonlinear programming~\cite{tanprasert2000analytical}, and force-directed methods~\cite{goplen2003efficient}.
In addition, {NTUPlace3-3D}~\cite{hsu2011tsv,hsu2013tsv} performs 3D analytical placement based on a bell-shaped~\cite{kahng2004implementation} smooth density considering TSV insertion, and mPL6-3D~\cite{luo2013analytical} utilizes a Huber-based local smoothing technique working with a Helmholtz-based global smoothing approach.
Based on mPL6-3D~\cite{luo2013analytical}, ART-3D~\cite{murali2022art} improves placement quality using reinforcement learning-based parameter tuning. The state-of-the-art analytical placement is the ePlace family~\cite{lu2013fftpl,lu2015eplace,lu2015eplacems,lu2016eplace} where the density constraint is modeled by an electrostatic field.
Lu et al. proposes a general 3D eDensity model in ePlace-3D~\cite{lu2016eplace} achieving analytically global smoothness along all dimensions in 3D domain. Remarkably, the ePlace family has achieved substantial success in wirelength-driven analytical placement, and their adoption of fast Fourier transform (FFT) for solving the 3D numerical solution has inspired quality enhancement~\cite{cheng2018replace} and GPU-accelerated ultra-fast implementations~\cite{lin2019dreamplace,liu2022xplace}.

Recently, Chen et al.~\cite{chen2023analytical} have proposed a 3D analytical placement algorithm to optimize wirelength considering F2F-bonded 3D ICs with multiple manufacturing technologies. They devise a multi-technologies weighted-average (MTWA) wirelength model using sigmoid-based functions for pin offset transition, and establish their framework based on ePlace-3D~\cite{lu2016eplace}. 
A 2D analytical placement, considering the accurate wirelength, is employed after the 3D global placement to further refine the solution.

\subsection{Analytical Placement}
\label{subsec:prelim-analytical-place}
The non-differentiability of the net wirelength in~\Cref{def:net-wirelength} and the modeling of legality constraints lead to difficulties in numerical optimization. Typically, we apply analytical placement approaches in global placement to find provably good cell locations with little overlap allowed, legalize all nodes to satisfy legality constraints in the legalization step, and refine locations in the detailed placement step.

Removing the utilization constraints in~\Cref{eq:high-level-formulation}, the analytical global placement problem is formulated as a numerical optimization
\begin{equation}
  \min_{\bm{v}}\sum_{e\in E}W_e(\bm{v})+\lambda D(\bm{v}),
  \label{eq:analytical-placement}
\end{equation}
where $\bm{v}$ indicates the node location variables, $W_e(\cdot)$ is the net wirelength model of net $e\in E$, $D(\cdot)$ is the density model of the entire placement region, and $\lambda$ is the density weight introduced as the Lagrangian multiplier of the density constraint.
The wirelength model $W_e(\cdot)$ in~\Cref{eq:analytical-placement} is usually a differentiable approximation~\cite{naylor2001non,hsu2011tsv,hsu2013tsv,liao2023on} to net HPWL.
\begin{definition}[3D HPWL]
  \label{def:3d-hpwl}
  Given node positions $\bm{x},\bm{y},\bm{z}$, the 3D HPWL of any net $e\in E$ is given by
  \begin{equation}
    W_e(\bm{x},\bm{y},\bm{z})=p_{e}(\bm{x})+p_{e}(\bm{y})+\alpha p_{e}(\bm{z}),
    \label{eq:3d-hpwl-def}
  \end{equation}
  where $p_e(\bm{u})=\max_{c_i\in e}u_i-\min_{c_i\in e}u_i$ denote the range or peak-to-peak function that evaluates the difference of maximum minus minimum in a net, and $\alpha\geq0$ is a weight factor.
\end{definition}
$p_e(\cdot)$ denotes partial HPWL along one axis. In real applications, it is approximated by a differentiable model, \emph{e.g.} the weighted-average~\cite{hsu2013tsv} model given a smoothing parameter $\gamma>0$,
\begin{equation}
  p_{e,\text{WA}}(\bm{u})=\frac{\sum_{c_i\in e}u_i\mathrm{e}^{\frac{1}{\gamma}u_i}}{\sum_{c_i\in e}\mathrm{e}^{\frac{1}{\gamma}u_i}}-\frac{\sum_{c_i\in e}u_i\mathrm{e}^{-\frac{1}{\gamma}u_i}}{\sum_{c_i\in e}\mathrm{e}^{-\frac{1}{\gamma}u_i}}.
  \label{eq:wa-model-approx}
\end{equation}
Other differentiable models~\cite{naylor2001non,liao2023on} are also applicable. Note that the $z$-dimension is usually defined manually, as tiers are discretely distributed in 3D scenarios. The corresponding weight factor $\alpha\geq0$ is determined in accordance with specific objectives in real applications.

The state-of-the-art density model $D(\cdot)$ is the {eDensity} family~\cite{lu2013fftpl,lu2015eplace,lu2015eplacems,lu2016eplace} based on electrostatics field, where every node $c_i\in V$ is modeled by an electric charge. We implement eDensity-3D~\cite{lu2016eplace} as our density model with GPU acceleration in the proposed framework.

The optimization formulation in~\Cref{eq:analytical-placement} is general and thus can be applied in both 2D and 3D analytical global placement. In conventional 2D cases, the variable $\bm{v}=(\bm{x},\bm{y})$ is optimized to find planar cell coordinates~\cite{lu2015eplace,lin2019dreamplace,liu2022xplace}. In 3D cases, the framework is well-established in {ePlace-3D}~\cite{lu2016eplace} where the $z$-direction coordinates is considered to optimize $\bm{v}=(\bm{x},\bm{y},\bm{z})$.

\section{Problem Formulation}
\label{sec:problem-formulation}

\subsection{Problem Statement}
\label{subsec:problem-statement}
In this paper, we focus on the 3D placement problem with die-to-die (D2D) connections, specified in the ICCAD 2022 Contest~\cite{hu20222022}. The general requirement is to partition the given standard cells into two dies with different technologies, create vertical interconnections named hybrid bonding terminals (HBTs) for split nets, and determine the locations of all nodes including standard cells and HBTs so that the following constraints are satisfied:
\begin{itemize}
\item\textbf{Utilization Constraints}. The utilization requirements of the top die and the bottom die are provided separately, leading to different area upper bound for two dies.
\item\textbf{Technology Constraints}. The cells may be fabricated using different technologies on different dies, \emph{i.e.}, the cell characteristic, cell height, cell width, and the cell layout would be different.
\item\textbf{Vertical Interconnection Constraints}. For any net $e$ split to two dies, an HBT should be created to connect pins on the top die and the bottom die. All HBTs share the same size.
\item\textbf{Legality Constraints}. All standard cells on both dies should be placed without overlap and aligned to rows and sites. HBTs should be placed to satisfy the spacing constraint, \emph{i.e.}, the distance between each pair of HBTs and the distance to boundaries are lower bounded.
\end{itemize}

The objective of this 3D placement problem is the total wirelength of all nets in the given design defined in~\Cref{def:net-wirelength}. In short, we focus on minimizing the sum of HPWL on the two dies. The center points of the HBTs are included in the HPWL calculation for each die. We will give rigorous mathematical formulations in~\Cref{subsec:problem-formulation}.

\subsection{Problem Formulation}
\label{subsec:problem-formulation}
Consider a netlist $(V,E)$ where node set $V=\{c_1,\cdots,c_n\}$ and net set $E=\{e_1,\cdots,e_m\}$. A partition is determined by a 0-1 vector $\bm{\delta}\in\mathbb{Z}_2^n=\{0,1\}^n$, where $\delta_i=0$ indicates that cell $c_i\in V$ is placed on the bottom die, otherwise top die. In the 3D placement with D2D vertical connections, the partition determines the total number of hybrid bonding terminals. In this section, we use $\bm{x},\bm{y}$ to represent both node coordinates and corresponding pin coordinates ignoring pin offsets for simplicity.

\begin{definition}[Net Cut Indicator]
  The cut indicator of a net $e\in E$ is a function of partition $\bm{\delta}\in\{0,1\}^n$ defined by
  \begin{equation}
    C_e(\bm{\delta})=\max_{c_i\in e}\delta_i-\min_{c_i\in e}\delta_i.
  \end{equation}
  It is also a binary value in $\{0,1\}$. If there exist two nodes incident to net $e$ placed on two different dies, the cut $C_e(\bm{\delta})=1$, otherwise it is 0.
\end{definition}

Given a partition $\bm{\delta}\in\{0,1\}^n$, if a net $e\in E$ is a split net, \emph{i.e.}, $C_e(\bm{\delta})=1$, a hybrid bonding terminal (HBT) should be inserted for this net as a vertical connection. Otherwise, all nodes incident to $e\in E$ are placed on either the top or the bottom die. Different from TSVs and MIVs going through silicon substrates, HBTs do not require silicon space. If we have $C_{e_i}(\bm{\delta})=1$ for a net $e_i\in E$, one and only one HBT $t_i$ should be assigned to $e_i$ accordingly, otherwise $t_i$ will be discarded. 
We denote the set of HBTs by $T=\{t_1,\cdots,t_m\}$ with planar coordinates $\bm{x}',\bm{y}'$.
%Since we treat HBTs as special nodes, we denote $c_{n+j}=t_{j}$ for $j=1,\cdots,m$ and similarly $(x_{n+j},y_{n+j})=(x_j',y_j')$ for simplicity. 

Denote the top and bottom partial nets by $e^{+}(\bm{\delta})=\{c_i\in e:\delta_i=1\}$ and $e^{-}(\bm{\delta})=\{c_i\in e:\delta_i=0\}$, respectively. Correspondingly, the complete nets on top and bottom dies are $\tilde{e}_i^{+}=e_i^{+}\cup\{t_i\}$ and $\tilde{e}_i^{-}=e_i^{-}\cup\{t_i\}$, respectively, including HBTs. The die-to-die (D2D) wirelength~\cite{hu20222022} of net $e\in E$ is defined as follows.
%If net $e_i\in E$ is not split, \emph{i.e.}, $C_{e_i}(\bm{\delta})=0$, either $e_i^{+}$ or $e_i^{-}$ is empty. For example, if all nodes connected by $e$ are on the top die, we have $e^{+}=e$ and $e^{-}=\varnothing$, otherwise we have $e^{+}=\varnothing$ and $e^{-}=e$. Accordingly, the HBT $t_i$ will be discarded and thus the augmented net $\tilde{e}_i$ is exactly $e_i$ itself.

 \begin{definition}[D2D Net Wirelength]
  \label{def:net-wirelength}
  Given partition $\bm{\delta}$, the die-to-die (D2D) wirelength of net $e$ is defined by $W_e=W_{\tilde{e}_i^{+}}+W_{\tilde{e}_i^{-}}$. More specifically, we have
  \begin{equation}
    \begin{aligned}
      W_{\tilde{e}_i^{+}}&=\max_{c_j\in\tilde{e}_i^{+}}x_j-\min_{c_j\in\tilde{e}_i^{+}}x_j+\max_{c_j\in\tilde{e}_i^{+}}y_j-\min_{c_j\in\tilde{e}_i^{+}}y_j,\\
      W_{\tilde{e}_i^{-}}&=\max_{c_j\in\tilde{e}_i^{-}}x_j-\min_{c_j\in\tilde{e}_i^{-}}x_j+\max_{c_j\in\tilde{e}_i^{-}}y_j-\min_{c_j\in\tilde{e}_i^{-}}y_j.
    \end{aligned}
    \label{eq:exact-wirelength}
  \end{equation}
  If $C_{e_i}(\bm{\delta})=0$, it degrades to the ordinary net HPWL without HBT considered.
\end{definition}

The D2D net wirelength in~\Cref{def:net-wirelength} simply sums up the half-perimeter wirelength on two dies, demonstrating equivalence to $p_{\tilde{e}_i^{+}}(\bm{x})+p_{\tilde{e}_i^{-}}(\bm{x})+p_{\tilde{e}_i^{+}}(\bm{y})+p_{\tilde{e}_i^{-}}(\bm{y})$.
Since the center point of HBT $t_i$ is included, $W_e$ is a function of node locations $\bm{x},\bm{y}$, HBT locations $\bm{x}',\bm{y}'$, and partition $\bm{\delta}$.
Our problem is formulated as follows.
\begin{equation}
  \begin{array}{cl}
    \displaystyle\min_{\bm{x},\bm{y},\bm{\delta},\bm{x}',\bm{y}'}&\displaystyle\sum_{e\in E}W_e(\bm{x},\bm{y},\bm{\delta},\bm{x}',\bm{y}')\\[2\jot]
    \mathrm{s.t.}&\displaystyle\sum_{i=1}^n\delta_ia_i^{+}\leq a_{\text{req}}^{+},\\
    &\displaystyle\sum_{i=1}^n(1-\delta_i)a_i^{-}\leq a_{\text{req}}^{-},\\
    &\text{legality constraints,}
  \end{array}
  \label{eq:high-level-formulation}
\end{equation}
where $a_i^{+},a_i^{-}$ stand for the node area of $c_i$ on the top and bottom die, respectively. The area requirements are set to $a_{\text{req}}^{+},a_{\text{req}}^{-}$ correspondingly. 
Besides of the legality constraints of standard cells on both dies, all HBTs have a specific legality rule that the distance between each other is lower bounded. 
It worth mentioning that HBTs are on the top-most metal layer and thus would not occupy any placement resources on both dies.

\section{Overall Placement Flow}
\label{sec:overall-placement-flow}
\begin{figure}[t]
  \centering
  \resizebox{\linewidth}{!}{
    \begin{tikzpicture}[
  box/.style={rectangle,rounded corners=3pt,fill=none,
    inner sep=1.5mm,minimum width=5cm,line width=1pt,align=center},
  innerbox/.style={box,draw=slategray},
  stepbox/.style={box,densely dashed,
    fill=none,draw=slategray,inner sep=2mm},
  seg/.style={rounded corners=3pt,line width=1pt,slategray},
  card/.style={rounded corners=3pt,line width=1pt,fill=white,
    draw=slategray,inner sep=2mm,node distance=.35cm,
    minimum width=3.6cm},
  document/.style={line width=1pt,shape=tape,
    draw=slategray!70,fill=white,minimum width=2cm,
    tape bend top=none},
  multidocument/.style={doc,double copy shadow},
  lseg/.style={seg,slategray!70,line width=3pt,
    -{Triangle[width=7pt,length=7pt]}},
  node distance=1mm,
]
  \pgfdeclarelayer{mid1}
  \pgfdeclarelayer{mid2}
  \pgfsetlayers{background,mid2,mid1,main}
  \node[innerbox,draw=slategray!60] (init) {Initialize Node Positions};
  \coordinate[above=2mm of init] (beforeinit);
  \node[inner sep=0,below=.8cm of init.south] (grads-anchor) {};
  \node[inner sep=0,left=2mm of grads-anchor] (grads) {};
  \node[innerbox,right=.5mm of grads,draw=slategray!70,minimum width=2cm,fill=slategray!5] (density-grad) {%
    \textbf{3D Electrostatics}\\\textbf{Density Gradient}};
  \node[innerbox,left=.5mm of grads,draw=slategray!70,minimum width=2cm,fill=slategray!5] (wl-grad) {%
    \textbf{Bistratal WL}\\\textbf{Gradient}};
  \node[box,fill=none,draw,densely dashed,draw=slategray!70,fit={(density-grad)(wl-grad)}] (all-grad) {};
  \node[innerbox,below=.8cm of grads-anchor,fill=slategray!10,draw=slategray!80] (update-pos) {Update Node Positions};
  \node[innerbox,below=of update-pos,fill=slategray!15,draw=slategray!90] (update-param) {Update Parameters};
  \node[innerbox,below=of update-param,fill=slategray!20] (update-attr) {Update Node Attributes};
  \node[innerbox,diamond,aspect=2,inner sep=.5mm,
    below=3mm of update-attr,fill=slategray!25] (check) {Converge?};
  \node[below=2mm of check] (aftercheck) {};
  \coordinate (grad-east) at ($(all-grad.east)+(.5,0)$);
  \path let\p1=(grad-east),\p2=(check) in coordinate (mid-point)
  at ($(\x1,\y1)!.5!(\x1,\y2)$);
  \draw[seg] (check)-|(mid-point);
  \draw[seg,-{Triangle[width=5pt,length=5pt]}]
  (mid-point)|-(all-grad);
  \draw[seg] (update-attr)--(check);
  \node[label=90:{No}] at (check.east) {};
  \node[label=-30:{Yes}] at (check.south) {};
  \begin{pgfonlayer}{mid1}
    \node[stepbox,fit={(beforeinit)(all-grad)(aftercheck)(mid-point)}]
    (global-place) {};
  \end{pgfonlayer}
  \node[label=45:\parbox{5em}{\textbf{3D Global}\\\textbf{Placement}}]
  at (global-place.south west) {};
  \path let\p1=(check),\p2=(global-place.south) in
  coordinate (gp-end) at ($(\x1,\y2)+(0,.1)$);
  \path let\p1=(init),\p2=(global-place.north) in
  coordinate (gp-beg) at ($(\x1,\y2)-(0,.1)$);
  \draw[seg,-{Triangle[width=5pt,length=5pt]},slategray] (check)--(gp-end);
  \draw[seg,-{Triangle[width=5pt,length=5pt]}] (gp-beg)--(init);

  % The right part
  \node[card,fill=slategray!20,draw=slategray,line width=1.5pt,
    above right=1.4cm and .6cm of global-place.east] (gp-card) {%
    \textbf{3D Global Placement}};
  \begin{pgfonlayer}{mid1}
    \fill[fill=slategray!10] (gp-card.north west)--(global-place.north east)
    --(global-place.south east)--(gp-card.south west)--cycle;
  \end{pgfonlayer}
  \node[document,above=6mm of gp-card] (input) {\textbf{Input}};
  \node[card,below=4mm of gp-card] (hbt-card) {HBT Assignment};
  \node[card,below=4mm of hbt-card] (lg-card) {Legalization};
  \node[card,below=4mm of lg-card] (dp-card) {Detailed Placement};
  \node[document,below=6mm of dp-card] (output) {\textbf{Output}};
  \begin{pgfonlayer}{mid1}
    \draw[lseg] (input.center)--(gp-card);
    \draw[lseg] (gp-card)--(hbt-card);
    \draw[lseg] (hbt-card)--(lg-card);
    \draw[lseg] (lg-card)--(dp-card);
    \draw[lseg] (dp-card)--(output);
  \end{pgfonlayer}
  
\end{tikzpicture}
  }
  \caption{The Overall Placement Flow of Our Framework.}
  \label{fig:overall-flow}
\end{figure}
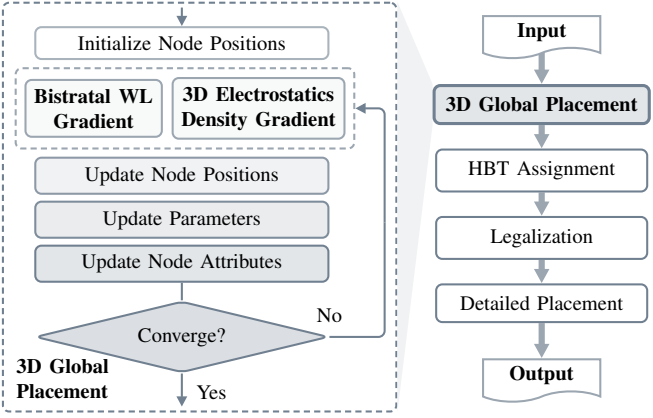
The overall placement flow of our proposed framework is illustrated in~\Cref{fig:overall-flow}. We adopt a 3D analytical global placement to find node locations with three dimensions. After global placement, we assign HBTs and legalize all nodes including HBTs. At last, we perform detailed placement on each die to further refine the solution. The optimized circuit placement results will be output after detailed placement. Note that we do not apply 2D placement after 3D global placement and HBT assignment, as we are confident enough of our proposed 3D global placement which effectively handles partitioning and planar placement together.

\subsection{Global Placement}
\label{subsec:gp}
In 3D circuit placement, we assign coordinates $\bm{x},\bm{y},\bm{z}\in\mathbb{R}^n$ to all nodes in the design.
Given the top die $[x_{\min}^{+},x_{\max}^{+}]\times[y_{\min}^{+},y_{\max}^{+}]$ and the bottom die $[x_{\min}^{-},x_{\max}^{-}]\times[y_{\min}^{-},y_{\max}^{-}]$, we have to make a necessary and realistic assumption that they differ very little so that the entire placement region is well-defined and the 3D placement framework makes sense under this scenario.
\begin{assumption}
  \label{as:die-area}
  The die sizes of two dies are almost the same. Specifically, we have die width $x_{\min}^{+}=x_{\min}^{-}=0$, $x_{\max}^{+}=x_{\max}^{-}$, and die height $y_{\min}^{+}=y_{\min}^{-}=0$, $\left|\frac{y_{\max}^{+}}{y_{\max}^{-}}-1\right|<\epsilon$, where $\epsilon>0$ is a small tolerance.
\end{assumption}
Under~\Cref{as:die-area}, our 3D global placement region is set to a cuboid $\Omega=[0,x_{\max}^{+}]\times[0,y_{\max}^{+}]\times[0,z_{\max}]$ by default, with a properly determined depth $z_{\max}$. For each node $c_i\in V$, along with its width and height provided by the input files, it will also be assigned a unified depth $d$.

Different from the 2D cases, the partition values $\bm{\delta}$ are restricted to take very discrete values in 3D placement to determine node partition. More specifically, $\bm{\delta}$ must be constrained to take binary values in $\{0,1\}^n$ in our placement problem, described in~\Cref{subsec:problem-formulation}, so that each node $c_i\in V$ has an assigned partition indicator. We equally split the placement cuboid $\Omega$ into two parts by the plane $z=\frac{1}{2}z_{\max}$, each of which represents a die:
\begin{equation}
  \begin{aligned}
    \Omega^{+}&=[0,x_{\max}^{+}]\times[0,y_{\max}^{+}]\times\left[\frac{z_{\max}}{2},z_{\max}\right]\\
    \Omega^{-}&=[0,x_{\max}^{+}]\times[0,y_{\max}^{+}]\times\left[0,\frac{z_{\max}}{2}\right].
  \end{aligned}
  \label{eq:placement-region}
\end{equation}
The unified node depth is $d=\frac{1}{2}z_{\max}$. Ideally, we expect every node $c_i\in V$ to be placed inside either the top part $\Omega^{+}$ or the bottom part $\Omega^{-}$ at the end of 3D global placement. Note that every node should not be placed out of boundary, therefore $z_i$, which stands for the corner point coordinate of node $c_i$, should take values within interval $[0,\frac{1}{2}z_{\max}]$. We determine the \emph{tentative} node partition $\bm{\delta}$ as a function of $z$ coordinates $P(\bm{z})$, by rounding the normalized value $\frac{2}{z_{\max}}\bm{z}$ at every iteration, \emph{i.e.}, we have
\begin{equation}
  \delta_i=\left\lceil\frac{2z_i}{z_{\max}}-\frac{1}{2}\right\rceil,
  \label{eq:partition-rounding}
\end{equation}
for every $c_i\in V$.

\begin{figure}
  % in tikz drawing
\newcommand\drawcuboid[5][]{%
  \draw[#1] ($#2+(#3,0,0)$)-- ++(0,#4,0)-- ++(-#3,0,0)-- ++(0,-#4,0)-- cycle;
  \draw[#1] ($#2+(#3,0,#5)$)-- ++(0,#4,0)-- ++(-#3,0,0)-- ++(0,-#4,0)-- cycle;
  \draw[#1] #2-- ++(0,0,#5);
  \draw[#1] ($#2+(#3,0,0)$)-- ++(0,0,#5);
  \draw[#1] ($#2+(#3,#4,0)$)-- ++(0,0,#5);
  \draw[#1] ($#2+(0,#4,0)$)-- ++(0,0,#5);
}
\newcommand\drawsolidcuboidback[6][]{
  \filldraw[box,fill=#2!50,draw=none,#1] #3-- ++(#4,0,0)-- ++(0,#5,0)-- ++(-#4,0,0)-- cycle;
  \filldraw[box,fill=#2!30,draw=none,#1] #3-- ++(0,#5,0)-- ++(0,0,#6)-- ++(0,-#5,0)-- cycle;  
  \filldraw[box,fill=#2!30,draw=none,#1] ($#3+(#4,#5,#6)$)-- ++(0,0,-#6)-- ++(-#4,0,0)-- ++(0,0,#6)-- cycle;
  \draw[box,#2,#1] #3-- ++(0,#5,0)-- ++(#4,0,0);
  \draw[box,#2,#1] ($#3+(0,#5,0)$)-- ++(0,0,#6);
}
\newcommand\drawsolidcuboidfront[6][]{
  \filldraw[box,fill=#2!50,draw=none,#1] ($#3+(0,0,#6)$)-- ++(0,#5,0)-- ++(#4,0,0)-- ++(0,-#5,0)-- cycle;
  \filldraw[box,fill=#2!30,draw=none,#1] #3-- ++(#4,0,0)-- ++(0,0,#6)-- ++(-#4,0,0)-- cycle;
  \filldraw[box,fill=#2!30,draw=none,#1] ($#3+(#4,0,0)$)-- ++(0,#5,0)-- ++(0,0,#6)-- ++(0,-#5,0)-- cycle;
  \draw[box,#2,#1] #3-- ++(#4,0,0)-- ++(0,#5,0)-- ++(0,0,#6)-- ++(-#4,0,0)-- ++(0,-#5,0)-- cycle;
  \draw[box,#2,#1] ($#3+(#4,0,0)$)-- ++(0,0,#6)-- ++(0,#5,0);
  \draw[box,#2,#1] ($#3+(#4,0,#6)$)-- ++(-#4,0,0);
}
\newcommand\drawsolidcuboid[6][]{%
  \drawsolidcuboidback[#1]{#2}{#3}{#4}{#5}{#6}
  \drawsolidcuboidfront[#1]{#2}{#3}{#4}{#5}{#6}
}
\begin{subfigure}[]{.49\linewidth}
  \centering
  \begin{tikzpicture}[
      line join=round,label distance=-1mm,
      box/.style={opacity=.7}
    ]
    \pgfdeclarelayer{fg}
    \pgfsetlayers{main,fg}
    \begin{axis}[
        grid=both,scale=.48,minor tick num=1,
        axis line style={draw=none},
        view={20}{20},line width=.8pt,
        zmin=0,zmax=2,z buffer=sort,
        xmin=0,xmax=10,
        ymin=0,ymax=10,
        xticklabels=\empty,
        yticklabels=\empty,
        zticklabels={,,$\frac{z_{\max}}{2}$,$z_{\max}$},
        %colormap={lblue}{rgb=(0,0.43,0.8),rgb=(0.9,0.9,0.85)},
        colormap={gray}{rgb=(0.46,0.46,0.46),rgb=(0.9,0.9,0.85)}
      ]
      \draw[gray] (0,0,0)--(0,10,0)--(10,10,0);
      \draw[gray] (0,10,0)--(0,10,2);
      \drawsolidcuboid{deepblue}{(2,3,0.2)}{2}{3}{0.8} % bottom box 1
      \drawsolidcuboid{rosybrown}{(6.5,4,0.65)}{2.2}{2.0}{0.35} % bottom box 2
      \addplot3[data cs=cart,surf,domain=0:10,samples=2,opacity=0.5]{1};
      \drawsolidcuboid{deepblue}{(2,3,1.0)}{2}{3}{0.2} % top box 1
      \drawsolidcuboid{rosybrown}{(6.5,4,1.0)}{2.2}{2.0}{0.65} % bottom box 2
      \node[label={-90:$c_i$}] at (3,4.5,0.2) {};
      \node[label={90:$c_j$}] at (7.6,5,1.65) {};
      % have to manually set the axes on the top most layer
      \begin{pgfonlayer}{fg}
        \draw[line width=.8pt] (0,0,0)--(0,0,2)--(0,10,2)--(10,10,2)--(10,10,0)--(10,0,0)--cycle;
      \end{pgfonlayer}
    \end{axis}
  \end{tikzpicture}
  \caption{}
  \label{subfig:partition}
\end{subfigure}
\begin{subfigure}[]{.49\linewidth}
  \centering
  \begin{tikzpicture}[
    line join=round,label distance=-1mm,
    box/.style={opacity=.7},
    elliparc/.style args={#1:#2:#3}{insert path={++(#1:#3) arc (#1:#2:#3)}}
  ]
    \pgfdeclarelayer{fg}
    \pgfsetlayers{main,fg}
    \begin{axis}[
        grid=both,scale=.48,minor tick num=1,
        axis line style={draw=none},
        view={20}{20},line width=.8pt,
        zmin=0,zmax=2,z buffer=sort,
        xmin=0,xmax=10,
        ymin=0,ymax=10,
        xticklabels=\empty,
        yticklabels=\empty,
        zticklabels={,,$\frac{z_{\max}}{2}$,$z_{\max}$},
        %colormap={lblue}{rgb=(0,0.43,0.8),rgb=(0.9,0.9,0.85)},
        colormap={gray}{rgb=(0.46,0.46,0.46),rgb=(0.9,0.9,0.85)}
      ]
      \draw[gray] (0,0,0)--(0,10,0)--(10,10,0);
      \draw[gray] (0,10,0)--(0,10,2);
      \drawsolidcuboidback{deepblue}{(2,3,0.0)}{3}{4.5}{1.0} % bottom box 1
      %\draw[-latex] (3,4.5,1.0)--(3,4.5,0.5);
      \drawsolidcuboidfront{deepblue}{(2,3,0.0)}{3}{4.5}{1.0} % bottom box 1
      \draw[canvas is xz plane at y=4.5,darkgray,
        opacity=.7,-latex,line width=1pt] (5,1) [elliparc=0:-90:2 and 0.5];
      \addplot3[data cs=cart,surf,domain=0:10,samples=2,opacity=0.5]{1};
      \drawsolidcuboidback[dashed]{rosybrown}{(2,3,1.0)}{2}{3}{1.0} % top box 1
      %\draw[] (3,4.5,1.5)--(3,4.5,1.0);
      \drawsolidcuboidfront[dashed]{rosybrown}{(2,3,1.0)}{2}{3}{1.0} % top box 1
      \draw[canvas is xz plane at y=4.5,darkgray,
        opacity=.7,line width=1pt] (5,1) [elliparc=90:0:2 and 0.5];
      \node[label={0:$c_i$}] at (4,6,1.8) {};
      \node[label={0:$c_i$}] at (5,3,0) {};
      % have to manually set the axes on the top most layer
      \begin{pgfonlayer}{fg}
        \draw[line width=.8pt] (0,0,0)--(0,0,2)--(0,10,2)--(10,10,2)--(10,10,0)--(10,0,0)--cycle;
      \end{pgfonlayer}
    \end{axis}
  \end{tikzpicture}
  \caption{}
  \label{subfig:update-attributes}
\end{subfigure}
\let\drawcuboid\undefined
\let\drawsolidcuboid\undefined
  \caption{
    The partition mapping $P(\bm{z}):[0,z_{\max}]\rightarrow\{0,1\}$ and the update of node attributes for heterogeneous technologies.
    \subref{subfig:partition}~At one iteration during 3D global placement, node $c_i$ has \emph{tentative} partition $\delta_i=0$ indicating the bottom die, while node $c_j$ with $\delta_j=1$ is assigned to the top die.
    \subref{subfig:update-attributes}~The node size and pin offset values of node $c_i\in V$ will change if moved to the other die.
  }
\end{figure}
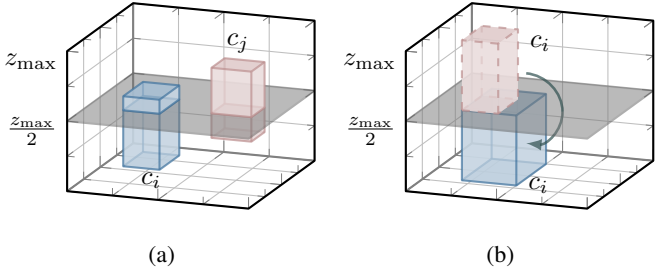

An example of partition mapping $\bm{\delta}=P(\bm{z})$ is depicted in~\Cref{subfig:partition}.
Node $c_i$ is partitioned to the bottom die, \emph{i.e.}, $\delta_i=0$ as its corner coordinate $z_i<\frac{1}{4}z_{\max}$.
The other node $c_j$ in~\Cref{subfig:partition} is partitioned to the top die, \emph{i.e.},
$\delta_i=1$ as its corner coordinate $z_j>\frac{1}{4}z_{\max}$.
The exact value of the cuboid depth $z_{\max}$ should be determined properly to avoid ill-condition in numerical optimization. We will discuss it in~\Cref{sec:results}.
\vskip .5em%

\textbf{Heterogeneous Technologies}.
Different from ordinary analytical placement, we have to face a challenge of heterogeneous technologies that the node attributes including node sizes and pin offset values are different on the two dies.

Assume that each node $c_i\in V$ has width $w_i^+$ and height $h_i^+$ on the top die and $w_i^{-},h_i^{-}$ on the bottom die. At each iteration of 3D global placement, we should determine the exact node size for every node according to tentative partition $\bm{\delta}=P(\bm{z})$. More specifically, if the tentative partition $\delta_i=1$, $w_i^{+},h_i^{+}$ will be adopted for node $c_i$, otherwise it will use $w_i^{-},h_i^{-}$. In other words, the planar node size for node $c_i\in V$ is calculated as
\begin{equation}
  \begin{aligned}
    w_i&=\delta_iw_i^{+}+(1-\delta_i)w_i^{-},\\
    h_i&=\delta_ih_i^{+}+(1-\delta_i)h_i^{-}
  \end{aligned}
  \label{eq:node-size-tech}
\end{equation}
where the tentative partition $\delta_i$ determined by~\Cref{eq:partition-rounding} is a binary value. The node depth remains $d=\frac{1}{2}z_{\max}$ in the entire process of 3D global placement.

In addition to the node size, we also have two sets of pin offset values, although they are ignored for simplicity in previous wirelength notations. Denote all pins by $P=\{p_1,\cdots,p_l\}$, and $P_{i}$ is the set of all pins on the node $c_i\in V$. Now, let $\bm{x}_{\text{offset}},\bm{y}_{\text{offset}},\bm{z}_{\text{offset}}\in\mathbb{R}^l$ be the pin offset vectors on three dimensions. For any $p_j\in P_i$, we have 
\begin{equation}
  \begin{aligned}
    x_{\text{offset},j}&=\delta_ix_{\text{offset},j}^{+}+(1-\delta_i)x_{\text{offset},j}^{-},\\
    y_{\text{offset},j}&=\delta_iy_{\text{offset},j}^{+}+(1-\delta_i)y_{\text{offset},j}^{-},
  \end{aligned}
  \label{eq:pin-offset-tech}
\end{equation}
and $z_{\text{offset},j}=\frac{1}{4}z_{\max}$ is fixed. In other words, the pin offset values of every pin is determined by the tentative partition of the node it belongs to. Besides, we have a fact that, for any $p_j\in P_i$, node $c_i$'s tentative partition $\delta_i=1$ if and only if pin $p_j$ is on the top part $\Omega^{+}$: $z_i+z_{\text{offset},j}\geq\frac{z_{\max}}{2}$.

In accordance with~\Cref{eq:node-size-tech} and~\Cref{eq:pin-offset-tech}, we update the \emph{node attributes} including node size and pin offset at every iteration during 3D global placement. An example of updating node attributes is illustrated in~\Cref{subfig:update-attributes} where node $c_i$ is moved from $z_i=\frac{1}{2}z_{\max}$ to $z_i=0$.
\vskip .5em%

\textbf{Bistratal Wirelength with Numerical Differentiation Approximation}.
Traditional 3D wirelength models, \emph{e.g.} the 3D HPWL model in~\Cref{def:3d-hpwl} handle $z$-direction wirelength and corresponding gradients differently to tackle the numerical inconsistency of coordinates along three axes, reflecting on a weight $\alpha$ while relaxing $\bm{z}$ to be continuous. However, these wirelength models are inaccurate to estimate the exact D2D wirelength.

Since the wirelength model is critical to the overall numerical optimization, we propose a \emph{bistratal wirelength model} with numerical differentiation approximation and will discuss the details in~\Cref{sec:wirelength-model}. Note that we do \emph{NOT} insert HBTs during global placement.
\vskip .5em%

\textbf{Electrostatics-Based 3D Density}. As mentioned in~\Cref{subsec:prelim-analytical-place}, {eDensity}~\cite{lu2015eplace} is the state-of-the-art academic density model which analogizes every node $c_i$ to a positive electric charge $q_i$. It expects an electric equilibrium so that movable objects can be evened out to reduce the overall node overlap. Extending the density model in~\cite{lu2015eplace}, {ePlace-3D}~\cite{lu2016eplace} computes the potential map by solving the 3D Possion's equation under Neumann boundary condition,
\begin{equation}
  \begin{array}{rl}
    \Delta\phi=-\rho,&\text{in }\Omega\\
    \hat{\bm{n}}\cdot\nabla\phi=0,&\text{on }\partial\Omega,
  \end{array}
  \label{eq:3d-poisson-equation}
\end{equation}
where $\rho=\rho(x,y,z)$ is the current density map in placement region $\Omega=[0,x_{\max}]\times[0,y_{\max}]\times[0,z_{\max}]$ computed using node locations. The second line in~\Cref{eq:3d-poisson-equation} is the boundary condition specifying that the electric force on the boundary is zero.

Suppose the placement region $\Omega$ is uniformly decomposed into $N_x\times N_y\times N_z$ grids, the solution to~\Cref{eq:3d-poisson-equation} under constraint $\int_{\Omega}\phi\,\mathrm{d}\Omega=0$ is given by
\begin{equation}
  \phi=\sum_{j,k,l}\frac{a_{jkl}}{\omega_j^2+\omega_k^2+\omega_l^2}\cos(\omega_jx)\cos(\omega_ky)\cos(\omega_lz),
  \label{eq:poisson-sol-potential}
\end{equation}
where the tuple $(\omega_j,\omega_k,\omega_l)=(\frac{j\pi}{x_{\max}},\frac{k\pi}{y_{\max}},\frac{l\pi}{z_{\max}})$ stands for frequency indices. The density coefficients $a_{jkl}$ is defined by
\begin{equation}
  a_{jkl}=\frac{1}{N}\sum_{x,y,z}\rho\cos(\omega_jx)\cos(\omega_ky)\cos(\omega_lz).
  \label{eq:density-coef}
\end{equation}
where the denominator $N=N_xN_yN_z$ denotes the total number of bins. Note that the DC component of density map $\rho$ has been removed, \emph{i.e.}, $\int_{\Omega}\rho\,\mathrm{d}\Omega=0$ is satisfied by removing $a_{000}=\frac{1}{N}\sum_{x,y,z}\rho(x,y,z)$ which equals to the average density of all bins. The electric field $\bm{E}(x,y,z)=(E_x,E_y,E_z)$ can be directly derived from~\Cref{eq:poisson-sol-potential} by taking partial derivatives of $\phi$,
\begin{equation}
  \begin{aligned}
    E_x&=\sum_{j,k,l}\frac{a_{jkl}\omega_j}{\omega_j^2+\omega_k^2+\omega_l^2}\sin(\omega_jx)\cos(\omega_ky)\cos(\omega_lz),\\
    E_y&=\sum_{j,k,l}\frac{a_{jkl}\omega_k}{\omega_j^2+\omega_k^2+\omega_l^2}\cos(\omega_jx)\sin(\omega_ky)\cos(\omega_lz),\\
    E_z&=\sum_{j,k,l}\frac{a_{jkl}\omega_l}{\omega_j^2+\omega_k^2+\omega_l^2}\cos(\omega_jx)\cos(\omega_ky)\sin(\omega_lz),
  \end{aligned}
  \label{eq:possion-sol-force}
\end{equation}
\Cref{eq:poisson-sol-potential} and~\Cref{eq:possion-sol-force} are well-established in~\cite{lu2016eplace}, demonstrating that these spectral equations can be solved efficiently using FFT with $O(N\log{N})$ time complexity.

Different from the general scenarios in~\cite{lu2016eplace} where they may have multiple tiers, we only have two dies in our specific problem. To help the 3D electrostatic filed even out the standard cells to different dies, the node depth is set to $d=\frac{1}{2}z_{\max}$ by default, as mentioned above. Through the numerical optimization of 3D global placement, standard cells are expected to be roughly distributed within either $\Omega^{+}$ or $\Omega^{-}$, so that the tentative partition $\bm{\delta}=P(\bm{z})$ does not introduce significant wirelength degradation after 3D global placement.

\subsection{HBT Assignment}
During 3D global placement, we do \emph{NOT} insert HBTs as any HBT is allowed to have overlap with standard cells. After 3D global placement, we first obtain a partition $\bm{\delta}=P(\bm{z})\in\mathbb{Z}_2^n$ according to~\Cref{eq:partition-rounding}. 
The convergence of global placement implies a very low overflow indicating that $z_i$ should be close to either 0 or $\frac{1}{2}z_{\max}$ to determine the partition solution. Since the partition $\bm{\delta}$ and $\bm{x},\bm{y}$ is already determined, we proceed to the 2D scenario with the top die layout and the bottom die layout. Every split net $e$ should be assigned precisely one HBT.
\begin{figure}[t]
  \centering
  \subfloat[]{
  \label{subfig:optimal-hbt-a}
  \begin{tikzpicture}[scale=.6,line width=1pt]
    \filldraw[gray!10,draw=none] (-1.5,-1.5) rectangle++ (6,6);
    \fill[deepblue,fill opacity=.5] (0,0) rectangle++ (2,2);
    \fill[rosybrown,fill opacity=.5] (1,1) rectangle++ (2,2);
    \draw[draw=deepblue!50!rosybrown]
    (1,1) rectangle++ (1,1);
    \node[label=-90:{$B_e^{-}$}] at (0,0) {};
    \node[label=90:{$B_e^{+}$}] at (3,3) {};
    \node[label={[label distance=-2mm]-45:{$B_{e,\text{HBT}}^*$}}] at (2,1) {};
  \end{tikzpicture}
}\hspace{2em}
\subfloat[]{
  \label{subfig:optimal-hbt-b}
  \begin{tikzpicture}[scale=.6,line width=1pt]
    \filldraw[gray!10,draw=none] (-1.5,-1.5) rectangle++ (6,6);
    \fill[deepblue,fill opacity=.5] (0,-0.75) rectangle++ (2,2);
    \fill[rosybrown,fill opacity=.5] (1,1.75) rectangle++ (2,2);
    \draw[draw=deepblue!50!rosybrown]
    (1,1.25) rectangle++ (1,0.5);
    \node[label=180:{$B_e^{-}$}] at (0,-0.75) {};
    \node[label=0:{$B_e^{+}$}] at (3,3.75) {};
    \node[label={[label distance=-2mm]-45:{$B_{e,\text{HBT}}^*$}}] at (2,1.25) {};
  \end{tikzpicture}
}\\[1em]
\subfloat[]{
  \label{subfig:optimal-hbt-c}
  \begin{tikzpicture}[scale=.6,line width=1pt]
    \filldraw[gray!10,draw=none] (-1.5,-1.5) rectangle++ (6,6);
    \fill[deepblue,fill opacity=.5] (-0.75,0) rectangle++ (2,2);
    \fill[rosybrown,fill opacity=.5] (1.75,1) rectangle++ (2,2);
    \draw[draw=deepblue!50!rosybrown]
    (1.25,1) rectangle++ (0.5,1);
    \node[label=-90:{$B_e^{-}$}] at (-0.75,0) {};
    \node[label=90:{$B_e^{+}$}] at (3.75,3) {};
    \node[label={[label distance=-1mm]-45:{$B_{e,\text{HBT}}^*$}}] at (1.25,1) {};
  \end{tikzpicture}
}\hspace{2em}
\subfloat[]{
  \label{subfig:optimal-hbt-d}
  \begin{tikzpicture}[scale=.6,line width=1pt]
    \filldraw[gray!10,draw=none] (-1.5,-1.5) rectangle++ (6,6);
    \fill[deepblue,fill opacity=.5] (-0.75,-0.75) rectangle++ (2,2);
    \fill[rosybrown,fill opacity=.5] (1.75,1.75) rectangle++ (2,2);
    \draw[draw=deepblue!50!rosybrown]
    (1.25,1.25) rectangle++ (0.5,0.5);
    \node[label=90:{$B_e^{-}$}] at (-0.75,1.25) {};
    \node[label=-90:{$B_e^{+}$}] at (3.75,1.75) {};
    \node[label={[label distance=-1mm]-45:{$B_{e,\text{HBT}}^*$}}] at (1.25,1.25) {};
  \end{tikzpicture}
}
  \caption{The optimal region $B_{e,\text{HBT}}^*$ of an HBT for a split net $e\in E$ with cut $C_e(\bm{\delta})=1$ under several different scenarios.
  \subref{subfig:optimal-hbt-a}~The top net bounding box $B_e^{+}$ and the bottom net bounding box $B_e^{-}$ overlap on both the $x$ dimension and the $y$ dimension.
  \subref{subfig:optimal-hbt-b}~$B_e^{+}$ and $B_e^{-}$ overlap only on the $x$ dimension.
  \subref{subfig:optimal-hbt-c}~$B_e^{+}$ and $B_e^{-}$ overlap only on the $y$ dimension.
  \subref{subfig:optimal-hbt-d}~$B_e^{+}$ and $B_e^{-}$ have no overlap on both two dimensions.}
  \label{fig:optimal-region}
\end{figure}
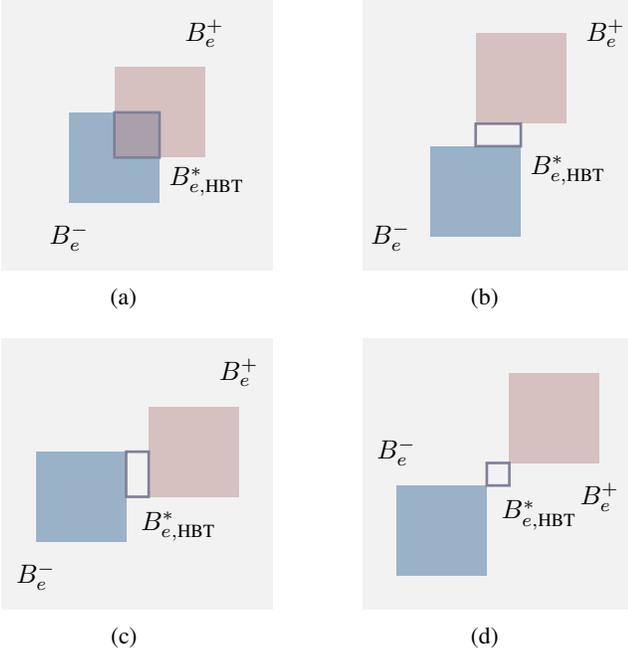

Consider a split net $e\in E$. Ignoring pin offset values for simplicity, define $x$-dimension coordinates $x_{\text{low}}^{+}=\min_{c_i\in e^{+}}x_i$, $x_{\text{high}}^{+}=\max_{c_i\in e^{+}}x_i$ and vertical coordinates $y_{\text{low}}^{+}, y_{\text{high}}^{+}$ for the top partial net $e^{+}$, and similarly define corresponding variables for the bottom partial net $e^{-}$.
Then, we denote the bounding box of partial nets $e^{+}$ and $e^{-}$ by
\begin{equation}
  \begin{aligned}
    B_e^{+}&=[x_{\text{low}}^{+},x_{\text{high}}^{+}]\times[y_{\text{low}}^{+},y_{\text{high}}^{+}],\\
    B_e^{-}&=[x_{\text{low}}^{-},x_{\text{high}}^{-}]\times[y_{\text{low}}^{-},y_{\text{high}}^{-}],
  \end{aligned}
  \label{eq:partial-net-box}
\end{equation}
respectively.

After 3D global placement, the $\bm{x},\bm{y}$ coordinates and partition $\bm{\delta}=P(\bm{z})$ of nodes are already determined, and thus $B_e^+$ and $B_e^-$ are determined for every split net $e$. As illustrated in~\Cref{fig:optimal-region}, for any split net $e$, its HBT has a specific \emph{optimal region}, \emph{i.e.}, the net wirelength $W_e$ is minimized only when its HBT is placed within this optimal region.
\begin{theorem}
  \label{thm:optimal-region-hbt}
  For a split net $e\in E$, the optimal region of its HBT is defined by $B_{e,\text{HBT}}^*=[x_{\text{low}}',x_{\text{high}}']\times[y_{\text{low}}',y_{\text{high}}']$ where
  \begin{equation}
    \begin{aligned}
      x_{\text{low}}'&=\makebox[\widthof{$\max$}][c]{$\min$}\left\{\max\left\{x_{\text{low}}^{+},x_{\text{low}}^{-}\right\},
      \min\left\{x_{\text{high}}^{+},x_{\text{high}}^{-}\right\}\right\},\\
      x_{\text{high}}'&=\max\left\{\max\left\{x_{\text{low}}^{+},x_{\text{low}}^{-}\right\},
      \min\left\{x_{\text{high}}^{+},x_{\text{high}}^{-}\right\}\right\},
    \end{aligned}
  \end{equation}
  and $y_{\text{low}}',y_{\text{high}}'$ are defined similarly.
  Equivalently, coordinates $x_{\text{low}}',x_{\text{high}}'$ are the two median numbers of $x_{\text{low}}^{+},x_{\text{low}}^{-},x_{\text{high}}^{+},x_{\text{high}}^{-}$ and the same for $y_{\text{low}}',y_{\text{high}}'$.
\end{theorem}

\Cref{thm:optimal-region-hbt} enlightens us that the total net wirelength will be minimized when every split net $e$ has its HBT placed within the optimal region $B_{e,\text{HBT}}^*$. Therefore, we intuitively assign an HBT $t(e)\in T$ for each split net such that the center point of $t$ locates exactly at the center point of $B_{e,\text{HBT}}^*$.

Note that after this \emph{HBT assignment} step, it is likely that HBTs may overlap with each other, requiring a subsequent legalization process. 
To control the total number of HBTs and mitigate potential wirelength degradation caused by legalization, we carefully regulate the weight $\alpha$ in the objective function described in~\Cref{def:3d-hpwl}.
This enables us to mitigate wirelength degradation while minimizing the number of HBTs.

\subsection{Legalization}
\label{subsec:legal}
After the partitioning $\bm{\delta}=P(\bm{z})$ and the HBT assignment, the mission of 3D global placement is completed. The rest is to legalize all nodes including HBTs and further refine the solution from 2D perspective. We legalize the standard cells on the top die and the bottom die separately with Tetris~\cite{hill2002method} and Abacus~\cite{spindler2008abacus}. The HBTs are legalized similarly by treating them as ordinary standard cells with a specific terminal size.

Note that in our problem definition, HBTs share the same square size $w'\times w'$ and every pair of HBTs must satisfy the spacing constraint that the distance of boundaries should be no less than $s'$. Hence, we pad every HBT to a square with size $w'+s'$ and legalize them as ordinary standard cells with row height $w'+s'$.

\subsection{Detailed Placement}
\label{subsec:dp}
We further improve the total wirelength by applying ABCDPlace~\cite{lin2020abcdplace} with several techniques including global swap~\cite{pan2005efficient,popovych2014density}, independent set matching~\cite{chen2008ntuplace3}, and local reordering~\cite{pan2005efficient,chen2008ntuplace3} die by die. When we are performing detailed placement on one die, all other nodes on the other die and HBTs remain fixed. After the detailed placement of two dies, the optimal regions of HBTs may get affected. Therefore, we can continue to map HBTs to their updated optimal regions, followed by a new round of HBT legalization and detailed placement. 
While this process can be iterated infinitely, we find that only the initial few rounds yield significant benefits. 
Therefore, we perform one additional round of this process during the detailed placement.

\section{Bistratal Wirelength Model}
\label{sec:wirelength-model}
The analytical wirelength model is critical to the numerical optimization of~\Cref{eq:analytical-placement} in this problem. Previous works~\cite{hsu2011tsv,hsu2013tsv,lu2016eplace} use the 3D HPWL model defined in~\Cref{def:3d-hpwl} with the peak-to-peak function to describe the net wirelength. Chen et al.~\cite{chen2023analytical} propose MTWA model to consider heterogeneous technologies, but it is still based on 3D HPWL without considering the D2D wirelength. 
Note that $p_e(\bm{z})$ roughly reflects the cut size of net $e$ and does not contribute to the planar net wirelength. 
The plain HPWL $\tilde{W}_e$ is defined as follows such that $W_e(\bm{x},\bm{y},\bm{z})=\tilde{W}_e(\bm{x},\bm{y})+\alpha p_e(\bm{z})$.
\begin{definition}[Plain HPWL]
  \label{def:plain-hpwl}
  Given node positions $\bm{x},\bm{y}$, the plain HPWL of any net $e\in E$ is given by
  \begin{equation}
    \tilde{W}_e(\bm{x},\bm{y})=\max_{c_i\in e}x_i-\min_{c_i\in e}x_i+\max_{c_i\in e}y_i-\min_{c_i\in e}y_i,
    \label{eq:plain-hpwl-def}
  \end{equation}
  which does not care node position $\bm{z}$ at all.
\end{definition}
Obviously,~\Cref{eq:plain-hpwl-def} in the above definition is equivalent to the separable representation $\tilde{W}_e(\bm{x},\bm{y})=p_e(\bm{x})+p_e(\bm{y})$ using the peak-to-peak function defined in~\Cref{eq:3d-hpwl-def}.

Unfortunately, 3D HPWL model in~\Cref{def:3d-hpwl} based on the plain HPWL is \underline{\emph{inaccurate}} as the exact wirelength defined in~\Cref{def:net-wirelength} and~\Cref{eq:exact-wirelength} sums up the HPWL on the top die and bottom die.
~\Cref{eq:plain-hpwl-def} only considers the entire bounding box with the top die and the bottom die together, neglecting the pin partition and the potential presence of HBTs. 
Additionally, the conventional 3D HPWL wirelength model is \emph{NOT} able to capture the wirelength variation resulting from different node partition.

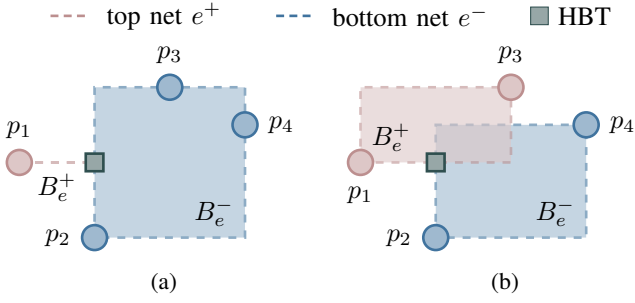
\begin{figure}[t]
  \centering
  \begin{tikzpicture}[
  net/.style={line width=1pt,densely dashed},
  hbt/.style={rectangle,fill=darkgray!50,draw=darkgray}
]
  \draw[net,rosybrown] (0,0)--(.5,0);
  \node[label=0:{top net $e^{+}$}] at (.5,0) {};
  \draw[net,deepblue] (3,0)--(3.5,0);
  \node[label=0:{bottom net $e^{-}$}] at (3.5,0) {};
  \node[hbt,label=0:{HBT}] (hbt) at (6.5,0) {};
  \end{tikzpicture}\\
\begin{subfigure}[]{.49\linewidth}
  \begin{tikzpicture}[
      line width=1pt,
      pin/.style={circle,fill},
      hbt/.style={rectangle,fill=darkgray!50,draw=darkgray},
      btmpin/.style={pin,fill=deepblue!50,draw=deepblue},
      toppin/.style={pin,fill=rosybrown!50,draw=rosybrown},
      net/.style={line width=1pt,dashed,opacity=.6}
    ]
    \node[toppin,label=90:$p_1$] (p1) at (0,0) {};
    \node[btmpin,label=180:$p_2$] (p2) at (1,-1) {};
    \node[btmpin,label=90:$p_3$] (p3) at (2,1) {};
    \node[btmpin,label=0:$p_4$] (p4) at (3,0.5) {};
    \node[hbt] (hbt) at (1,0) {};
    \begin{pgfonlayer}{background}
      \filldraw[net,draw=deepblue,fill=deepblue!50] (p2) rectangle (3,1);
      \filldraw[net,draw=rosybrown,fill=rosybrown!50] (0,0) rectangle (1,0);
    \end{pgfonlayer}
    \node[inner sep=0,label=-90:$B_e^{+}$] at ($(p1)!.5!(hbt)$) {};
    \path let\p1=(p2.center),\p2=(p4.center) in
    node[inner sep=0,label=135:$B_e^{-}$] at ($(\x2,\y1)$) {};
  \end{tikzpicture}
  \caption{}
  \label{subfig:hpwl-example-a}
\end{subfigure}
\begin{subfigure}[]{.49\linewidth}
  \begin{tikzpicture}[
      line width=1pt,
      pin/.style={circle,fill},
      hbt/.style={rectangle,fill=darkgray!50,draw=darkgray},
      btmpin/.style={pin,fill=deepblue!50,draw=deepblue},
      toppin/.style={pin,fill=rosybrown!50,draw=rosybrown},
      net/.style={line width=1pt,dashed,opacity=.6}
    ]
    \node[toppin,label=-90:$p_1$] (p1) at (0,0) {};
    \node[btmpin,label=180:$p_2$] (p2) at (1,-1) {};
    \node[toppin,label=90:$p_3$] (p3) at (2,1) {};
    \node[btmpin,label=0:$p_4$] (p4) at (3,0.5) {};
    \node[hbt] (hbt) at (1,0) {};
    \begin{pgfonlayer}{background}
      \filldraw[net,draw=deepblue,fill=deepblue!50] (p2) rectangle (p4);
      \filldraw[net,draw=rosybrown,fill=rosybrown!50] (p1) rectangle (p3);
    \end{pgfonlayer}
    \node[inner sep=0,label=45:$B_e^{+}$] at (p1) {};
    \path let\p1=(p2.center),\p2=(p4.center) in
    node[inner sep=0,label=135:$B_e^{-}$] at ($(\x2,\y1)$) {};
  \end{tikzpicture}
  \caption{}
  \label{subfig:hpwl-example-b}
\end{subfigure}
  \caption{An example where changing partition of one pin does not affect the net bounding box but increases the exact net wirelength.
    \subref{subfig:hpwl-example-a}~The exact wirelength equals to the HPWL of the entire net.
    \subref{subfig:hpwl-example-b}~The exact wirelength is strictly larger than the HPWL of the entire net.
  }
  \label{fig:hpwl-partition-effect}
\end{figure}
Consider a net $e\in E$ connecting four pins $p_1,p_2,p_3,p_4$. Fix all planar locations of these pins and tentative partition of $p_2,p_3,p_4$.~\Cref{subfig:hpwl-example-a} shows $e^{+}$ and $e^{-}$ when $p_3$ is on the bottom die and the corresponding HBT is placed optimally. It is clear that the total wirelength of net $e$ is $W_e=\tilde{W}_{\tilde{e}^{+}}+\tilde{W}_{\tilde{e}^{-}}$ which exactly equals to the \emph{plain} HPWL of the entire net $e$. By contrast,~\Cref{subfig:hpwl-example-b} shows the case when $p_3$ is on the top die. The HBT with the same coordinates preserves optimality, but the true wirelength $W_e=\tilde{W}_{\tilde{e}^{+}}+\tilde{W}_{\tilde{e}^{-}}$ is larger than the \emph{plain} HPWL of net $e$.
\begin{theorem}
  \label{thm:lower-bound-exact-wl}
  Given any partition $\bm{\delta}$ and any net $e\in E$, let $p_e(\bm{u})=\max_{c_i\in e}u_i-\min_{c_i\in e}u_i$ be the peak-to-peak function defined in~\Cref{eq:3d-hpwl-def}.
  Then, we always have
  \begin{equation}
    p_{e}\leq \min_{x'}W_{e_x}(x')\leq 2p_e,
    \label{ineq:bound-hpwl-exact-wl-part}
  \end{equation}
  where $W_{e_x}(x')$ is the $x$-dimension part of the exact net wirelength defined in~\Cref{eq:exact-wirelength} with HBT coordinate $x'$ under tentative partition $\bm{\delta}$.
\end{theorem}
The equality of the left part of~\Cref{ineq:bound-hpwl-exact-wl-part} holds if and only if $B_e^{+}$ and $B_e^{-}$ defined in~\Cref{eq:partial-net-box} has no overlap on the $x$ dimension. The equality of the right part holds if and only if $B_e^{+}$ and $B_e^{-}$ are the same on the $x$ dimension. The conclusion on the $y$ dimension can be similarly established. We will give a more detailed representation of $\min_{x'}W_{e_x}(x')$ in~\Cref{thm:forward-pass-obj}.
\begin{corollary}
  \label{crl:lower-bound-exact-wl}
  Given any partition $\bm{\delta}$ and any net $e\in E$, let the ordinary plain HPWL be $\tilde{W}_e$ defined in~\Cref{def:plain-hpwl}.
  Then, we always have
  \begin{equation}
    \tilde{W}_{e}\leq \min_{x',y'}W_e(x',y')\leq 2\tilde{W}_e,
    \label{ineq:bound-hpwl-exact-wl}
  \end{equation}
  where $W_e(\bm{\delta},x',y')$ is the exact net wirelength defined in~\Cref{eq:exact-wirelength} with HBT coordinate $(x',y')$.
\end{corollary}
The equality of the left part of~\Cref{ineq:bound-hpwl-exact-wl} holds if and only if $B_e^{+}$ and $B_e^{-}$ defined in~\Cref{eq:partial-net-box} has no overlap on both $x$ and $y$ dimensions. The equality of the right part holds if and only if $B_e^{+}$ and $B_e^{-}$ are the same.
\Cref{crl:lower-bound-exact-wl} indicates that the HPWL model used in previous works~\cite{hsu2011tsv,hsu2013tsv,lu2016eplace} is just a lower bound of the exact bistratal wirelength in our problem. Apparently, optimizing $\tilde{W}_{e}$ does not necessarily benefit the exact wirelength as the error bound may get as large as the lower bound, according to~\Cref{ineq:bound-hpwl-exact-wl}. We will give a precise representation of $\min_{x',y'}W_e(x',y')$ for every net $e$ in~\Cref{thm:forward-pass-obj}.

We propose a novel \emph{bistratal} wirelength model that handles planar coordinates and partitioning together. Instead of optimizing $\tilde{W}_e$, we try to minimize $\min_{\bm{x}',\bm{y}'}W_e(\bm{x},\bm{y},\bm{\delta},\bm{x}',\bm{y}')$ at every iteration according to the tentative partition. Besides of the wirelength estimation, the computation of gradients is more critical to the numerical optimization process. In this section, we will discuss the proposed model theoretically in detail.

\subsection{Wirelength Objective}
\label{subsec:wirelength-objective}
In the forward pass of numerical optimization~\cite{lu2015eplace,lin2019dreamplace}, we calculate the exact or approximated wirelength. Different from 3D HPWL in~\Cref{def:3d-hpwl}, we must consider partition for more precise wirelength estimation.

According to~\Cref{crl:lower-bound-exact-wl}, we should approximate the exact wirelength $W_e$ defined in~\Cref{eq:exact-wirelength} as precisely as possible. Considering that HBTs are not inserted in the 3D global placement as the tentative partition $\bm{\delta}$ may vary at every iteration, we assume that each split net is assigned a dummy HBT placed within its optimal region according to the tentative partition. In other words, we target at optimizing $\min_{\bm{x}',\bm{y}'}W_e(\bm{x},\bm{y},\bm{\delta},\bm{x}',\bm{y}')$ where the tentative partition $\bm{\delta}=P(\bm{z})$ is updated at every iteration. The following theorem reveals the explicit representation of our wirelength forward computation without any HBT inserted.
\begin{theorem}
  \label{thm:forward-pass-obj}
  The minimal precise net wirelength on the $x$ dimension with respect to the HBT coordinate $x'$ of net $e$ is given by
  \begin{equation}
    \min_{x'}W_{e_x}(x')=\max\left\{p_e,p_{e^{+}}+p_{e^{-}}\right\},
    \label{eq:forward-pass-obj}
  \end{equation}
  as a function of node positions $(\bm{x})$ under partition $\bm{\delta}$, where the peak-to-peak function $p_e$ is defined by $p_e(\bm{u})=\max_{c_i\in e}u_i-\min_{c_i\in e}u_i$ for any $\bm{u}$.
\end{theorem}
\Cref{thm:forward-pass-obj} gives an accurate estimation of the minimum exact net wirelength on the $x$ dimension for split nets at every iteration during 3D global placement. Note that the right-hand side of~\Cref{eq:forward-pass-obj} also indicates the exact wirelength for any non-split net $e$ as either $p_{e^{+}}$ or $p_{e^{+}}$ is zero. The corresponding theorem on the $y$ dimension can be similarly established.

Given any partition $\bm{\delta}$ and any split net $e\in E$, let $B_e^{+}=[x_{\text{low}}^{+},x_{\text{high}}^{+}]\times[y_{\text{low}}^{+},y_{\text{high}}^{+}]$ and $B_e^{-}=[x_{\text{low}}^{-},x_{\text{high}}^{-}]\times[y_{\text{low}}^{-},y_{\text{high}}^{-}]$ be the bounding boxes of partial nets $e^{+},e^{-}$, respectively, defined in~\Cref{eq:partial-net-box}. Define
\begin{equation}
  \hspace{-.4em}
  {W_e}_x=\left\{
  \begin{array}{ll}
    \displaystyle\max_{c_i\in e}x_i-\min_{c_i\in e}x_i,&\text{if }x_{\text{high}}\leq x_{\text{low}},\\[2\jot]
    \displaystyle x_{\text{high}}^{+}-x_{\text{low}}^{+}+x_{\text{high}}^{-}-x_{\text{low}}^{-},&\text{otherwise}.
  \end{array}
  \right.
  \label{eq:sub-obj-horizontal}
\end{equation}
where $x_{\text{low}}=\max\{x_{\text{low}}^{+},x_{\text{low}}^{-}\},x_{\text{high}}=\min\{x_{\text{high}}^{+},x_{\text{high}}^{-}\}$, and similary define ${W_e}_y$.
Then, the minimal precise net wirelength considering both $x,y$ dimensions with respect to the HBT coordinates $(x',y')$ of net $e\in E$ defined in~\Cref{thm:forward-pass-obj} is equivalent to
\begin{equation}
  \min_{x',y'}W_e(x',y')={W_e}_x+{W_e}_y,
  \label{eq:forward-pass-obj-equiv}
\end{equation}

More intuitively,~\Cref{eq:forward-pass-obj-equiv} first checks whether the boxes $B_e^{+}$ and $B_e^{-}$ overlap. If they overlap on one dimension, we optimize the HPWL of the top partial net $e^{+}$ and the bottom partial net $e^{-}$ on this dimension separately, as we have
\begin{equation}
  \begin{aligned}
    x_{\text{high}}^{+}-x_{\text{low}}^{+}&=p_{e^{+}}(\bm{x})=\max_{c_i\in e^{+}}x_i-\min_{c_i\in e^{+}}x_i,\\
    x_{\text{high}}^{-}-x_{\text{low}}^{-}&=p_{e^{-}}(\bm{x})=\max_{c_i\in e^{-}}x_i-\min_{c_i\in e^{-}}x_i,
  \end{aligned}
\end{equation}
otherwise the target degrades to the ordinary HPWL function $\tilde{W}_e$ on this dimension.

For a non-split net $e\in E$ with $C_e(\bm{\delta})=0$, \emph{i.e.}, it is completely within either the top or the bottom die, we treat it as an ordinary 2D net and evaluate its ordinary plain wirelength $\tilde{W}_e$ with~\Cref{eq:plain-hpwl-def}. Then, we propose the \emph{bistratal wirelength} (BiHPWL) as follows.

%From~\Cref{thm:forward-pass-obj}, we directly use~\Cref{eq:forward-pass-obj} as the target objective of our numerical optimization, which is more accurate to the exact wirelength and more sensitive to partition.

\begin{definition}[Bistratal Wirelength]
  \label{def:bistratal-wl}
  Given 3D node position $(\bm{x},\bm{y},\bm{z})$, the bistratal half-perimeter wirelength of any net $e$ is defined as
  \begin{multline}
    W_{e,\text{Bi}}(\bm{x},\bm{y},\bm{z})=
    \max\left\{p_e(\bm{x}),p_{e^{+}}(\bm{x})+p_{e^{-}}(\bm{x})\right\}+\\
    \max\left\{p_e(\bm{y}),p_{e^{+}}(\bm{y})+p_{e^{-}}(\bm{y})\right\}
    \label{eq:bistratal-wl}
  \end{multline}
  where the peak-to-peak function $p_e(\cdot)$ is defined by $p_e(\bm{u})=\max_{c_i\in e}u_i-\min_{c_i\in e}u_i$ for any $\bm{u}$. The partial nets $e^{+}(\bm{\delta})$ and $e^{-}(\bm{\delta})$ are determined by the tentative partition $\bm{\delta}=P(\bm{z})$.
\end{definition}

\Cref{def:bistratal-wl} gives a much accurate wirelength estimation in our probelm. Combining the regularization of cut size, In our 3D global placement, we use
\begin{equation}
  W(\bm{x},\bm{y},\bm{z}):=\sum_{e\in E}W_{e,\text{Bi}}(\bm{x},\bm{y},\bm{z})+\alpha\sum_{e\in E}p_e(\bm{z})
  \label{eq:3d-wl-general-obj}
\end{equation}
as the wirelength objective where the bistratal net wirelength $W_{e,\text{Bi}}$ is defined by~\Cref{eq:bistratal-wl}. Note that $W_{e,\text{Bi}}$ is also a function of $\bm{z}$ as the tentative partition $\bm{\delta}$ at every iteration is determined by $\bm{z}$. The second term with $\alpha$ weight is integrated to limit the total number of HBTs as we always expect fewer HBTs if possible. Moreover, a large number of HBTs would degrade the solution quality after legalization.

Optimizing \Cref{eq:3d-wl-general-obj} resolves the issue that 3D HPWL approximates the true wirelength poorly when the top box $B_e^{+}$ and the bottom box $B_e^{-}$ overlap, illustrated in~\Cref{subfig:hpwl-example-b}. However, the objective in~\Cref{eq:3d-wl-general-obj}
is highly non-differentiable. Therefore, we should establish the gradient approximation in detail to enable numerical optimization of 3D global placement. In the following of this section, we will discuss the gradient computation including the subgradient approximation to the planar gradients and the finite difference approximation to the depth gradient.

\subsection{Gradient Computation}
\label{subsec:gradient-computation}
The optimization of~\Cref{eq:3d-wl-general-obj} itself is difficult as it is non-differentiable and even discontinuous with respect to $\bm{z}$. In this subsection, we will discuss our proposed strategy to find the ``\emph{gradients}'' that percept the objective change with respect to variables.

Since the representations in~\Cref{eq:sub-obj-horizontal} are always in a peak-to-peak form, we use the weighted-average model~\cite{hsu2011tsv,hsu2013tsv} in~\Cref{eq:wa-model-approx} to approximate them, so that $W_e$ in~\Cref{eq:forward-pass-obj} is \emph{differentiable} where $x_{\text{high}}\neq x_ {\text{low}}$ and $y_{\text{high}}\neq y_{\text{low}}$ when calculating gradients. It is straght-forward to derive the closed-form representation of gradients of the WA model\cite{hsu2011tsv} described in~\Cref{eq:wa-model-approx},
\begin{equation}
  \frac{\partial p_{e,\text{WA}}}{\partial u_i}=
  \frac{\mathrm{e}^{\frac{u_i}{\gamma}}\left(\gamma+u_i-S_{\max}\right)}{\gamma\sum_{c_i\in e}\mathrm{e}^{\frac{u_i}{\gamma}}}-
  \frac{\mathrm{e}^{-\frac{u_i}{\gamma}}\left(\gamma+S_{\min}-u_i\right)}{\gamma\sum_{c_i\in e}\mathrm{e}^{-\frac{u_i}{\gamma}}},
  \label{eq:grad-wa-p2p}
\end{equation}
where the smooth maximum $S_{\max}=S_{\max}(\bm{u})$ and the smooth minimum $S_{\min}=S_{\min}(\bm{u})$ are defined by
\begin{equation}
  S_{\max}=\frac{\sum_{c_i\in e}u_i\mathrm{e}^{\frac{1}{\gamma}u_i}}{\sum_{c_i\in e}\mathrm{e}^{\frac{1}{\gamma}u_i}},\;\;
  S_{\min}=\frac{\sum_{c_i\in e}u_i\mathrm{e}^{-\frac{1}{\gamma}u_i}}{\sum_{c_i\in e}\mathrm{e}^{-\frac{1}{\gamma}u_i}},
\end{equation}
such that $p_{e,\text{WA}}=S_{\max}-S_{\min}$. The variable $\bm{u}$ can be $\bm{x},\bm{y},\bm{z}$ to derive the detailed gradients of the smooth peak-to-peak on corresponding dimensions. More details of differentiable approximations are discussed in~\cite{naylor2001non,hsu2011tsv,hsu2013tsv,liao2023on}.
\vskip .5em%

\textbf{Adaptive Planar Gradients}. In the numerical optimization, we are supposed to derive the ``gradients'' of~\Cref{eq:3d-wl-general-obj} with respect to coordinates $\bm{x},\bm{y},\bm{z}$. The gradients w.r.t planar coordinates $\bm{x},\bm{y}$ guide the optimizer to find optimal placement on each die, while the gradients w.r.t $\bm{z}$ handle the partition correspondingly. It is clear that the planar gradients are determined by $\nabla_{\bm{x}}W_{e,\text{Bi}}$ and $\nabla_{\bm{y}}W_{e,\text{Bi}}$. Unfortunately, $W_{e,\text{Bi}}$ in~\Cref{eq:bistratal-wl} is non-differentiable, forcing us to consider \emph{subgradients} instead.

Without loss of generality, we focus on the $x$ dimension. Consider function set $\mathcal{F}=\{p_e,p_{e^{+}}+p_{e^{-}}\}$ for a given tentative partition, then the $x$-dimension part of wirelength $W_{e,\text{Bi}}$ is $W_{e_x,\text{Bi}}(\bm{x})=\max_{f\in\mathcal{F}}f(\bm{x})$. The corresponding active function set is
\begin{equation}
  \mathcal{I}(\bm{x})=\{f\in\mathcal{F}:f(\bm{x})=W_{e_x,\text{Bi}}(\bm{x})\}.
  \label{eq:active-set}
\end{equation}
According to the subgradient calculus rule, we know that the subdifferential of $W_{e,\text{Bi}}$ is a convex hull
\begin{equation}
  \partial W_{e_x,\text{Bi}}(\bm{x})=\mathrm{conv}\bigcup_{f\in\mathcal{I}(\bm{x})}\partial f(\bm{x}).
  \label{eq:subdiff-calculus}
\end{equation}
We expect to legitimately take one subgradient $g\in\partial W_{e_x,\text{Bi}}(\bm{x})$ for optimization.

A non-split net is trivial as $W_{e_x,\text{Bi}}(\bm{x})$ degrades to $p_e(\bm{x})$ directly. Consider a split net $e\in E$. When $B_e^{+}$ and $B_e^{-}$ have overlap on $x$ dimension, \emph{i.e.} $p_{e^{+}}(\bm{x})+p_{e^{-}}(\bm{x})>p_{e}(\bm{x})$, $p_{e^{+}}(\bm{x})+p_e^{-}(\bm{x})\in\mathcal{I}(\bm{x})$ is active in~\Cref{eq:active-set} and we have $W_{e_x,\text{Bi}}(\bm{x})=p_{e^{+}}(\bm{x})+p_{e^{-}}(\bm{x})$. According to~\Cref{eq:subdiff-calculus}, it is straight-forward to take any subgradient in $\partial p_{e^{+}}(\bm{x})+\partial p_{e^{-}}(\bm{x})$ for numerical optimization. Empirically, differentiable approximations of $p_e$ may be preferred to work with smooth optimizers, and thus we take $\nabla_{\bm{x}}p_{e^{+},\text{WA}}+\nabla_{\bm{x}}p_{e^{-},\text{WA}}$ as the ``gradient'' $\nabla_{\bm{x}}W_{e_x,\text{Bi}}$, where we leverage the weighted-average model $p_{e,\text{WA}}$~\cite{hsu2011tsv,hsu2013tsv} defined in~\Cref{eq:wa-model-approx}. When $B_e^{+}$ and $B_e^{-}$ do not overlap on $x$ dimension, \emph{i.e.}, $p_{e^{+}}(\bm{x})+p_{e^{-}}(\bm{x})<p_{e}(\bm{x})$, $p_e(\bm{x})$ is active in~\Cref{eq:active-set}, so we have $W_{e_x,\text{Bi}}(\bm{x})=p_e(\bm{x})$ and treat $e$ as a non-split net, then apply the approximation $p_{e,\text{WA}}$. When $\mathcal{I}(\bm{x})$ is not a singleton, \emph{i.e.}, $p_{e^{+}}(\bm{x})+p_{e^{-}}(\bm{x})=p_{e}(\bm{x})$, we can take any element in the convex hull in~\Cref{eq:subdiff-calculus}. Through this way, we define the ``gradient'' $\nabla W_{e,\text{Bi}}$.
\begin{definition}[Planar Gradient]
  \label{def:planar-grad}
  Consider the bistratal wirelength $W_{e,\text{Bi}}$. The planar gradient $\nabla_{\bm{x}}W_{e,\text{Bi}}=\bm{g}$ is defined as follows
  \begin{equation}
    \bm{g}=\left\{
    \begin{array}{ll}
      \nabla p_{e^{+},\text{WA}}+\nabla p_{e^{-},\text{WA}},&\text{if }p_{e^{+}}+p_{e^{-}}>p_{e},\\[2\jot]
      \nabla p_{e,\text{WA}},&\text{otherwise}.
    \end{array}
    \right.
    \label{eq:grad-adaptive-planar}
  \end{equation}
  which is an approximation of a subgradient, where the gradient of the weighted-average approximation $p_{e,\text{WA}}$ is given by~\Cref{eq:grad-wa-p2p}.
\end{definition}
The gradient $\nabla_{\bm{y}}W_{e,\text{Bi}}$ can be defined similarly. Note that we still use the term $\nabla W_{e,\text{Bi}}$ to denote such a subgradient approximation in~\Cref{eq:grad-adaptive-planar} although $W_{e,\text{Bi}}$ itself is non-differentiable.

We consider~\Cref{eq:grad-adaptive-planar} to be the \emph{adaptive} planar gradients w.r.t. $\bm{x},\bm{y}$ coordinates. The term ``adaptive'' is named after the overlap illustrated in~\Cref{fig:hpwl-partition-effect}. More specifically, we check whether $B_e^{+}$ and $B_e^{-}$ overlap on $x$ (and $y$) dimensions under tentative $\bm{\delta}$ for every net $e$ at every global placement iteration. If they overlap on the $x$ (or $y$) dimension, we have $p_{e^{+}}+p_{e^{-}}>p_e$ and use the first representation of $\nabla W_{e,\text{Bi}}$ in~\Cref{eq:grad-adaptive-planar} and the second otherwise.~\Cref{eq:grad-adaptive-planar} is applied in our numerical optimization during the 3D global placement. With no doubt, it takes into account the physical information of pin coordinates on both dies, making it much more accurate than the 3D HPWL model.
\vskip .5em%

\textbf{Finite Difference Approximation of Depth Gradients}.
In addition to the planar gradients w.r.t. $\bm{x}$ and $\bm{y}$, we are also supposed to derive how to correctly define ``gradients'' w.r.t. $\bm{z}$ which is far more tricky. Finding a way to optimize $\bm{z}$ is critical to the entire optimization, as it directly determines the quality of partition.

The density gradient $\nabla_{\bm{z}}D(\bm{x},\bm{y},\bm{z})$ drives placer to separate nodes with depth $\frac{1}{2}z_{\max}$ to be distributed on two dies so that we can obtain a valid partition at last, neglecting wirelength optimization. The gradient $\sum_e\nabla_{\bm{z}}p_{e,\text{WA}}(\bm{z})$ in~\Cref{eq:3d-wl-general-obj} with the weighted-average model~\cite{hsu2013tsv} tends to optimize the total cutsize of the design so that the total number of HBTs is limited, but there is no theoretical guarantee that a small cutsize would benefit the D2D wirelength. Hence, the most important task is to find how $W_{e,\text{Bi}}(\bm{x},\bm{y},\bm{z})$ gets affected by $\bm{z}$ to evaluate the quality of partitioning. Considering that $W_{e,\text{Bi}}(\bm{x},\bm{y},\bm{z})$ is even discontinuous with respect to $\bm{z}$, the gradient $\nabla_{\bm{z}}W_{e,\text{Bi}}$ does not exist at all. To tackle this problem, we leverage \emph{finite difference} to approximate the impact of $\bm{z}$ on the bistratal wirelength.

Finite difference~\cite{taylor1717methodus,boole1960treatise,jordan1956calculus,milne2000calculus} has been widely used in a large number of applications in numerical differentiation to approximate derivatives. We follow the definitions and notations in~\cite{milne2000calculus} and denote the \emph{difference quotient} by
\begin{equation}
  \fdiff_hf(x)=\frac{f(x+h)-f(x)}{h}
\end{equation}
using the N\"{o}rlund's operator $\fdiff\limits_{\scriptscriptstyle h}$~\cite{milne2000calculus,norlund1924vorlesungen} for any function $f$ on $\mathbb{R}$ and $x,h\in\mathbb{R}$. In the classical infinitesimal calculus, the first-order derivative of $f$ is defined by $\lim_{h\rightarrow0}\fdiff\limits_{\scriptscriptstyle h}f(x)$ if $f$ is differentiable. Both difference and derivative estimate how the function value would change with its variables, but derivative is in a continuous view while difference depends on the step size $h$.

Taking a net $e\in E$ and $c_i\in e$, consider the impact of $z_i$ to the bistratal wirelength $W_{e,\text{Bi}}$. For simplicity, we use $W_{e,\text{Bi}}(z_i)$ to represent the bistratal wirelength of net $e\in E$ as a function of $z_i$ and fix all other variables. Given step size $h$, the difference quotient of $W_{e,\text{Bi}}$ at $z_i$ is
\begin{equation}
  \fdiff_{h}W_{e,\text{Bi}}(z_i)=\frac{W_{e,\text{Bi}}(z_i+h)-W_{e,\text{Bi}}(z_i)}{h},
  \label{eq:differential-quotient}
\end{equation}
combining both the \emph{forward/advancing difference} ($h>0$) and the \emph{backward/receding difference} ($h<0$).
Since $W_{e,\text{Bi}}$ is discontinuous with respect to $z_i$, the limit $\lim_{h\rightarrow0}\fdiff\limits_{\scriptscriptstyle h}W_{e,\text{Bi}}(z_i)$ does not exist. However, we could consider~\Cref{eq:differential-quotient} with a large $h$ as we only have two dies. More specifically, we set $h=\frac{1}{4}z_{\max}$ if $\delta_i=P(z_i)=0$ and $h=-\frac{1}{4}z_{\max}$ otherwise, so that the difference quotient in~\Cref{eq:differential-quotient} will always be non-zero. Providing that $W_{e,\text{Bi}}(z_i)$ is a step function that only takes two possible values $W_{e,\text{Bi}}(0)$ and $W_{e,\text{Bi}}(\frac{1}{2}z_{\max})$,~\Cref{eq:differential-quotient} can be summarized as follows.
\begin{definition}[Finite Difference Approximation]
  \label{def:difference-approx}
  Consider the bistratal wirelength $W_{e,\text{Bi}}$. The finite difference approximation (FDA) $\nabla_{\bm{z}}W_{e,\text{Bi}}=\bm{g}$ is defined by
  \begin{equation}
    \bm{g}_i=\fdiff_{\scalebox{.75}{$\frac{1}{4}z_{\max}$}}W_{e,\text{Bi}}(z_i)=\frac{4}{z_{\max}}\left(W_{e,\text{Bi}}\left(\frac{z_{\max}}{2}\right)-W_{e,\text{Bi}}(0)\right),
    \label{eq:grad-finite-diff}
  \end{equation}
  where $z_{\max}$ is the total depth of our placement region, defined in~\Cref{eq:placement-region}.
\end{definition}
\Cref{eq:grad-finite-diff} is intuitive that it actually evaluates the wirelength change when moving a pin to the other die. It provides a local view of benefits we can obtain when changing node partition. Note that we still use the term $\nabla_{\bm{z}}W_{e,\text{Bi}}$ to denote the finite difference approximation in~\Cref{def:difference-approx}, although $W_{e,\text{Bi}}$ itself is non-differentiable.

From~\Cref{eq:grad-finite-diff}, any node $c_i\in V$ accumulates depth gradients $\nabla_{\bm{z}}W_{e,\text{Bi}}$ from all related nets $e$, therefore the finite difference approximation locally evaluates the impact of every node to the total circuit wirelength. We apply $\sum_{e}\nabla_{\bm{z}}W_{e,\text{Bi}}$ in~\Cref{eq:grad-finite-diff} with cutsize gradient $\sum_e\nabla_{\bm{z}}p_{e,\text{WA}}(\bm{z})$ and density gradient $\nabla_{\bm{z}}D(\bm{x},\bm{y},\bm{z})$ to numerical optimization in 3D global placement to obtain a good partition with an acceptable number of HBTs. Combining with the adaptive planar gradients in~\Cref{def:planar-grad}, we have defined the detailed gradient computation of the proposed bistratal wirelength model.
% utilization handling, preconditioning
\section{Experimental Results}
\label{sec:results}
\subsection{Experimental Setup}
We conducted experiments on ICCAD 2022 contest benchmark suits~\cite{hu20222022}. The detailed design statistics are shown in~\Cref{tab:benchmark-stat}. 
Most of the designs adopt heterogeneous technologies for the two dies, bringing a significant challenge to optimize the total wirelength. 
Movable macros are not included in the benchmark suits.

We implemented the proposed 3D analytical placement framework in \texttt{C++} and \texttt{CUDA} based on the open-source placer {DREAMPlace}~\cite{lin2019dreamplace}. 
All the experiments were performed on a Linux machine with 16 Intel Xeon Gold 6226R cores (2.90GHz), 1 GeForce RTX 3090Ti graphics card, and 24 GB of main memory.
We compared our framework with the state-of-the-art (SOTA) placers from the top-3 teams in ICCAD 2022 contest and recent work~\cite{chen2023analytical}, and the reported results were evaluated by the official evaluator provided by the contest.

% \todo{$z$ coordinate distribution after 3D GP.}

\begin{table}[tb!]
  \centering\small
  \caption{The statistics of the ICCAD 2022 Contest Benchmark Suites~\cite{hu20222022} where $u^{+},u^{-}$ stand for the utilization constraints on the top die and the bottom die, respectively.}
  \label{tab:benchmark-stat}
  \resizebox{\linewidth}{!}{
    \begin{tabular}{l|*{5}{r}c}
      \toprule
      \multicolumn{1}{c|}{\textbf{Bench.}}&
      \multicolumn{1}{c}{\#Nodes}&
      \multicolumn{1}{c}{\#Nets}&
      \multicolumn{1}{c}{\#Pins}&
      \multicolumn{1}{c}{$u^{+}$}&
      \multicolumn{1}{c}{$u^{-}$}&
      \multicolumn{1}{c}{Diff Tech}\\
      \midrule
      \texttt{case2}  &2735   &2644   &8118   &0.70 &0.75 & Yes\\
      \texttt{case2h} &2735   &2644   &8118   &0.79 &0.79 & No \\
      \texttt{case3}  &44764  &44360  &142246 &0.78 &0.78 & No \\
      \texttt{case3h} &44764  &44360  &142246 &0.68 &0.78 & Yes\\
      \texttt{case4}  &220845 &220071 &773551 &0.66 &0.70 & Yes\\
      \texttt{case4h} &220845 &220071 &773551 &0.66 &0.76 & Yes\\
      \bottomrule
    \end{tabular}
  }
\end{table}

\subsection{Comparison with SOTA Placers}
\begin{table*}[t]
  \centering
  \caption{The experimental results on the ICCAD 2022 Contest Benchmarks~\cite{hu20222022} compared to the top-3 winners and the SOTA analytical 3D placer~\cite{chen2023analytical}. \textbf{WL} indicates the \emph{exact} D2D wirelength evaluated by the provided official evaluator. \textbf{HBTs} represents the cut size, \emph{i.e.}, the total number of hybrid bonding terminals. \textbf{RT} (s) stands for the total runtime.}
  \label{tab:main-result}
  \resizebox{\linewidth}{!}
  {
    %\normalsize\footnotesize
    \begin{tabular}{l*{2}{|S[table-format=9.0]S[table-format=5.0]S[table-format=4.0]}*{1}{|S[table-format=9.0]S[table-format=5.0]S[table-format=3.0]}*{2}{|S[table-format=9.0]S[table-format=5.0]S[table-format=4.0]}}
      \toprule
      \multicolumn{1}{c|}{\multirow{2}{*}{\textbf{Bench.}}} &
      \multicolumn{3}{c|}{1st Place} &
      \multicolumn{3}{c|}{2nd Place} &
      \multicolumn{3}{c|}{3rd Place} &
      \multicolumn{3}{c|}{~\cite{chen2023analytical}} &
      \multicolumn{3}{c}{Ours} \\
      &WL &HBTs &RT &WL &HBTs &RT &WL &HBTs &RT &WL &HBTs &RT &WL &HBTs &RT \\\midrule
      \texttt{case2} &2072075 &1131 & 47 &2080647 &477 & 7 &2097487 & 163 & 5 & 2011447 & 784 & 33 &\ubold 1944656 &646 & 38 \\
      \texttt{case2h} &2555461 &1083 & 45 &2735158 &687 & 8 &2644791 & 151 & 5 & 2514597 & 891 & 32 &\ubold 2462553 &345 & 40 \\
      \texttt{case3} &30580336 &16820 & 342 &30969011 &11257 & 234 &33063568 &14788 & 68 & 30302643 & 8169 & 141 &\ubold 30062713 & 8017 & 92 \\
      \texttt{case3h} &27650329 &16414 & 224 &27756492 &8953 & 243 &28372567 &11211 & 63 & 27135602 & 7727 & 155 &\ubold 26727327 & 8887 & 93 \\
      \texttt{case4} &281315669 &84069 & 1324 &274026687 &51480 & 1675 &281378049 &46468 & 391 & 272327370 & 53264 & 1189 &\ubold 267400694 & 42763 & 135 \\
      \texttt{case4h} &301193374 &84728 & 1096 &308359159 &59896 & 2040 &307399565 &58860 & 427 & 296655075 & 49616 & 1190 &\ubold 289245472 & 47712 & 146 \\\midrule
      \multicolumn{1}{c|}{Avg.} & {1.041} & {2.096} & {4.300} & {1.057} & {1.267} & {5.320} & {1.072} & {1.019} & {1.249} & {1.021} & {1.328} & {3.639} & {1.000} & {1.000} & {1.000} \\
      \bottomrule
    \end{tabular}
  }
\end{table*}
\Cref{tab:main-result} shows the experimental results of the top-3 teams, SOTA
analytical 3D placer~\cite{chen2023analytical}, and our framework on the contest benchmark suites~\cite{hu20222022}.
We compared the exact D2D wirelength (WL), the total number of hybrid bonding terminals (HBTs), and runtime of each case with the baselines.
The wirelength is evaluated using the provided official evaluator from the benchmark suites.
For a fair comparison, we acquired their binary executable files and evaluated the end-to-end runtime of the baselines on our machine using their default settings.

As illustrated in~\Cref{tab:main-result}, our analytical 3D placement framework consistently obtained the best WL results for all the cases, demonstrating the significant advantage of our 3D placement paradigm with the dedicated bistratal wirelength model.
Compared to the top-3 teams, our placer achieved 4.1\%, 5.7\%, and 7.2\% shorter wirelength on average, respectively.

Thanks to the global optimization view of our 3D analytical approach, our placer utilized fewer HBTs and achieved better wirelength.
Our framework reduced 52.3\%, 21.1\%, 2.0\% number of HBTs on average compared to thetop-3 contest winners.
Our framework achieved up tp 49.2\% HBT number reduction than the first place on the large cases, making our framework more competitive to reduce the hybrid bonding terminal fabrication cost for large designs in real scenarios.
Leveraging the computation power of modern GPUs, our placer demonstrates better runtime scalability than the baselines, achieving 4.300$\times$ and 5.320$\times$ speedup over the first place and the second place for end-to-end placement, and achieving up to 2.925$\times$ speedup over the third place on the large cases.

We also compared our proposed framework with the SOTA analytical 3D placer~\cite{chen2023analytical} on the same ICCAD 2022 benchmarks~\cite{hu20222022}. Chen et al.~\cite{chen2023analytical} proposed an MTWA wirelength model based on 3D HPWL in~\Cref{def:3d-hpwl}, considering heterogeneous technologies with a weight factor $\alpha$ that correlates positively with net degrees. 
They aimed to guide the optimizer to split more low-degree nets for wirelength reduction, which resulted in notable improvements compared to the first-place winner.
However, their wirelength model in 3D analytical placement is still inaccurate and thus requires an additional 2D placement to refine node locations. 
Moreover, MTWA~\cite{chen2023analytical} is not directly partitioning-aware. 
The experimental results in~\Cref{tab:main-result} show that we achieve up to 3.4\% wirelength improvement and 2.1\% on average
compared to~\cite{chen2023analytical}. Remarkably, we are confident enough of our placement framework in numerical optimization, and thus do not require a 2D placement to further refine node locations after 3D placement.
In addition, we require 25\% fewer HBTs and can efficiently accomplish the placement task with GPU resources. 
In modern VLSI design, the performance on large cases is most critical. 
We considered two large cases containing more than 220K standard cells in the ICCAD 2022 Contest benchmark suites~\cite{hu20222022}. 
As shown in~\Cref{tab:main-result}, we significantly outperform the baseline by 1.8\% and 2.5\% wirelength improvement on the largest two cases \texttt{case4} and \texttt{case4h}, respectively, with more than 8$\times$ runtime acceleration, proving the scalability of our proposed analytical framework.

\subsection{3D Global Placement Analysis}
\begin{figure}[t]
  \centering
  \begin{subfigure}{.48\linewidth}
    \includegraphics[width=\linewidth]{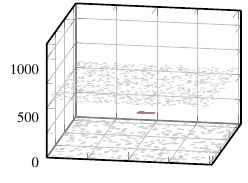}
    \caption{Iteration 0, WL 4.20$\times$\power{5},\\ Cut Size 1487, Overflow 0.96}
    \label{subfig:iter0}
  \end{subfigure}\hspace{.7em}
  \begin{subfigure}{.48\linewidth}
    \includegraphics[width=\linewidth]{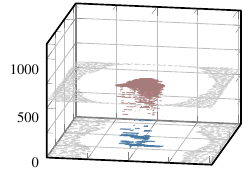}
    \caption{Iteration 800, WL 5.73$\times$\power{5},\\ Cut Size 60, Overflow 0.91}
    \label{subfig:iter800}
  \end{subfigure}\\[3\jot]
  \begin{subfigure}{.48\linewidth}
    \includegraphics[width=\linewidth]{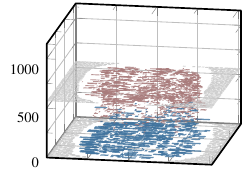}
    \caption{Iteration 1200, WL 1.48$\times$\power{6},\\ Cut Size 652, Overflow 0.47}
    \label{subfig:iter1200}
  \end{subfigure}\hspace{.7em}
  \begin{subfigure}{.48\linewidth}
    \includegraphics[width=\linewidth]{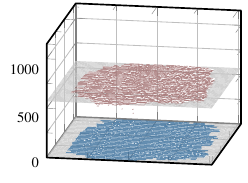}
    \caption{Iteration 1769, WL 1.71$\times$\power{6},\\ Cut Size 646, Overflow 0.07}
    \label{subfig:iter1769}
  \end{subfigure}
  \caption{
   3D global placement on \texttt{case2} with heterogeneous technologies. Fillers, nodes on the top die, and nodes on the bottom die are denoted by gray, brown, and blue rectangles, respectively. The node depth is omitted for better visualization. The nodes are initialized at the center point, and the fillers are randomly distributed on the two dies as shown in \subref{subfig:iter0}. The 3D density force combined with the wirelength force progressively drive all the nodes to the specific die, leading to a placement solution with almost perfect node partition as shown in \subref{subfig:iter1769}.
  }
  \label{fig:gp-iter}
\end{figure}
Our 3D analytical placement framework enables the simultaneous node partitioning and placement in the global placement stage, forming a larger solution space than previous separate partitioning and placement works~\cite{chang2016cascade2d,panth2017shrunk,ku2018compact,park2020pseudo}.
Unlike previous 3D analytical placer~\cite{lu2016eplace} targets on multiple tiers and leverages subsequent 2D placement to refine the placement solution, our framework assigns the nodes to exact two dies and place them in a single run.
Our 3D global placement is visualized in~\Cref{fig:gp-iter}.

In~\Cref{fig:gp-iter}, fillers, nodes on the top die, and nodes on the bottom die are denoted by gray, brown, and blue rectangles, respectively. The node depth is omitted for better visualization. All standard cells are randomly initialized around the center point of the design from a normal distribution, shown in~\Cref{subfig:iter0}.
Note that fillers are already inserted according to the given utilization requirements and uniformly initialized on two dies. During the 3D global placement, the optimizer tends to move nodes according to the gradients of wirelength (including cutsize with weight $\alpha$) and density.
The tentative partition $\bm{\delta}$ is updated at every intermediate iteration of global placement, shown in~\Cref{subfig:iter800} and~\Cref{subfig:iter1200}, until the convergence is detected. At last, the placer will find a 3D placement solution with optimized wirelength, shown in~\Cref{subfig:iter1769}.
When the convergence is attained, most standard cells $c_i$ with coordinate $z_i$ satisfying $|\frac{2z_i}{z_{\max}}-\delta_i|<\epsilon$ for a sufficiently small positive number $\epsilon>0$, implying that our framework is confident enough to partition every standard cell.
The 3D global placement produces a solution with overflow 0.07, shown in~\Cref{subfig:iter1769}, therefore we apply the tentative partition at the 1769th iteration as the final partition $\bm{\delta}$ and proceed to the later steps including legalization and detailed placement.

\subsection{Ablation Studies on Wirelength Models}
\begin{table*}[t]
  \centering
  \caption{The results of ablation studies on the ICCAD 2022 Contest Benchmarks~\cite{hu20222022} using different wirelength models with the same experimental settings. \textbf{WL} indicates the \emph{exact} D2D wirelength evaluated by the provided official evaluator. \textbf{HBTs} represents the cut size, \emph{i.e.}, the total number of hybrid bonding terminals. BiHPWL is the bistratal wirelength equipped with adaptive planar gradient in~\Cref{def:planar-grad}. FDA indicates finite difference approximation of depth gradients in~\Cref{def:difference-approx}.}
  \label{tab:ablation}
  %\resizebox{.8\linewidth}{!}
  {
    \footnotesize
    \begin{tabular}{l*{2}{|S[table-format=9.0]S[table-format=5.0]}*{2}{|S[table-format=9.0]S[table-format=5.0]}}
      \toprule
      \multicolumn{1}{c|}{\multirow{2}{*}{\textbf{Bench.}}} &
      \multicolumn{2}{c|}{Plain HPWL} &
      \multicolumn{2}{c|}{BiHPWL w/o. FDA} &
      \multicolumn{2}{c|}{Plain HPWL w/. FDA} &
      \multicolumn{2}{c}{BiHPWL w/. FDA} \\
      &WL &HBTs &WL &HBTs &WL &HBTs &WL &HBTs \\\midrule
      \texttt{case2} &2351813 &459 &2271554 &454 &2118450 &708 &\ubold 1944656 &646 \\
      \texttt{case2h} &2919815 &236 &2755549 &245 &3001905 &441 &\ubold 2462553 &345 \\
      \texttt{case3} &34776108 &4396 &33965431 &4547 &35577287 &8086 &\ubold 30062713 &8017 \\
      \texttt{case3h} &31093130 &4770 &30066866 &4781 &30748977 &8544 &\ubold 26727327 &8887 \\
      \texttt{case4} &309580785 &19339 &304667903 &23261 &288957440 &51369 &\ubold 267400694 &42763 \\
      \texttt{case4h} &330290736 &18971 &325610343 &22195 &324613980 &54942 &\ubold 289245272 &47712 \\\midrule
      \multicolumn{1}{c|}{Avg.} & {1.169} & {0.555} & {1.134} & {0.588} & {1.141} & {1.116} & {1.000} & {1.000} \\
      \bottomrule
    \end{tabular}
  }
\end{table*}
We evaluated the effectiveness of our proposed bistratal wirelength model by using different wirelength models in our framework on the ICCAD 2022 Contest Benchmarks~\cite{hu20222022}.
The detailed experimental results are shown in~\Cref{tab:ablation}.

Plain HPWL stands for the conventional HPWL model $\tilde{W}_e$ defined in~\Cref{def:plain-hpwl}. It is integrated in 3D HPWL adopted in many previous analytical placers~\cite{luo2013analytical,hsu2011tsv,hsu2013tsv,lu2016eplace}. This wirelength model is very classical and has been proved to be effective in analytical 3D placement. Notably, the gradients of differentiable approximations to $\tilde{W}_e$ w.r.t. $\bm{z}$ only focus on optimization on cutsize. Hence, it achieves the best results of cutsize, with only 55.5\% HBTs of ours, shown in~\Cref{tab:ablation}. However, the wirelength reported by the evaluator is 16.9\% larger than ours, as plain HPWL model could not comprehend the impact of partitioning on the exact D2D wirelength.
We now validate the effectiveness of the adaptive planar gradient defined in~\Cref{def:planar-grad} and the finite difference approximation (FDA) of depth gradients in~\Cref{def:difference-approx}.

BiHPWL in the second main column of~\Cref{tab:ablation} represents the bistratal wirelength in~\Cref{def:bistratal-wl} equipped with the adaptive planar gradient.
``BiHPWL model without FDA'' is equivalent to ``plain HPWL with adaptive planar gradient'' in terms of gradient computation.
As shown in~\Cref{tab:ablation}, the BiHPWL model without FDA achieves 3.5\% wirelength improvements on average with little degradation of cutsize, compared to plain HPWL. It is intuitively rational as the adaptive planar gradient tries to figure out when the plain HPWL is inaccurate compared to the exact D2D wirelength and switches a different strategy accordingly. However, it is still far inferior to the results with FDA, as the adaptive planar gradient in~\Cref{def:planar-grad} focuses on optimizations of planar coordinates $\bm{x},\bm{y}$ without comprehension of partitioning.

In the third main column of~\Cref{tab:ablation}, the plain HPWL is equipped with FDA, which means that we use $\tilde{W}_e$ to replace $W_{e,\text{Bi}}$ in~\Cref{def:difference-approx}. 
However, the plain HPWL $\tilde{W}_e$ is irrelevant to $\bm{z}$ and thus insensitive to different partitioning. 
Therefore, nonzero gradients occur only because of changes of node attributes given heterogeneous technologies, resulting in less than 3\% wirelength improvements with significant cutsize degradation.
By contrast, BiHPWL is evidently sensitive to partitioning, leading to 13.4\% wirelength improvement when FDA is enabled, 
as shown in the last main column in~\Cref{tab:ablation}.
Note that we utilize much more resources of vertical interconnects to optimize wirelength versus plain HPWL,
fully taking advantage of the benefits of F2F-bonded 3D ICs.
Meanwhile, our framework still significantly outperforms the first-place winner on cutsize while preserving advantages on wirelength.

\subsection{Runtime Breakdown}
\begin{figure}
  \centering
  {\begin{tikzpicture}[
    align=center,
    root/.style={
      circle,fill=white,
      text=gray,align=center,
      inner sep=1pt,          
    },
    scale=.75
  ]
  \small
  \definecolor{midwheat}{RGB}{196,178,143}
  % arctext from Andrew code with modifications:
  % Variables: 1: ID, 2:Style 3:box height 4: Radious 5:start-angl 6:end-angl 7:text {format along path}
  \def\arctext[#1][#2][#3](#4)(#5)(#6)#7{
    \draw[white,thick,line width=1pt,fill=#2]
    (#5:#4cm+#3) coordinate (above #1) arc (#5:#6:#4cm+#3)
    -- (#6:#4) coordinate (right #1)
    -- (#6:#4cm-#3) coordinate (below right #1) 
    arc (#6:#5:#4cm-#3) coordinate (below #1)
    -- (#5:#4) coordinate (left #1) -- cycle;
    \path[
      decoration={
        raise=-0.5ex, % Controls relavite text height position.
        text along path,
        text={#7},
        text align=center,
      }, decorate
    ] (#5:#4) arc (#5:#6:#4);
    \coordinate (#1 top) at ($(.5*#5+.5*#6:#4cm+#3)$);
    \coordinate (#1 center) at ($(.5*#5+.5*#6:#4cm)$);
  }
  \def\guidel(#1)#2#3#4#5#6{
    \draw[-,>=latex,color=#2,line width=1pt,
      shorten >=-2pt,shorten <=-2pt]
    let \p1=(#1),\n1={veclen(\x1,\y1)},\n2={atan2(\y1,\x1)}
    in (\n2:\n1)--(\n2:\n1+#3)--
    ({(\n1+#3)*cos(\n2)+#4-2*#5*#4},{(\n1+#3)*sin(\n2)});
    \path let \p1=(#1),\n1={veclen(\x1,\y1)},\n2={atan2(\y1,\x1)} in
    node[fill,circle,inner sep=1pt,minimum size=4pt] at (\n2:\n1-2pt) {};
    \path let \p1=(#1),\n1={veclen(\x1,\y1)},\n2={atan2(\y1,\x1)} in
    node[label={#5*180}:{#6}] at
    ({(\n1+#3)*cos(\n2)+#4-2*#5*#4},{(\n1+#3)*sin(\n2)}) {};
  }
  
  % Drawing the center
  % \node[root](root) at (0,0) {%
  %   {\normalsize Detailed}\\[3pt]%
  %   {\normalsize Placement}};
  % Drawing the Tex Arcs 
  \arctext[n1-fileio][midwheat][.5cm](2.0)(110)(100){};
  \arctext[n1-gp][rosybrown!80][.5cm](2.0)(100)(-194.628){};
  \arctext[n1-legalother][cyan!30][.5cm](2.0)(-194.628)(-202.828){};
  \arctext[n1-dp][deepblue!80][.5cm](2.0)(-202.828)(-250){};
  \guidel(n1-fileio center){black}{1.25cm}{4mm}{0}{File IO \& Others\\1.76\%}
  \guidel(n1-gp center){black}{1.25cm}{4mm}{0}{3D Global\\Placement\\82.59\%}
  \guidel(n1-legalother center){black}{1.25cm}{4mm}{1}{Legalization\\0.58\%}
  \guidel(n1-dp center){black}{1.25cm}{4mm}{1}{Detailed\\Placement\\15.07\%}
  \let\guidel\undefined
\end{tikzpicture}}
  \caption{The runtime breakdown of our proposed analytical 3D placement framework on the ICCAD 2022 Contest benchmark \texttt{case4h}~\cite{hu20222022}.}
  \label{fig:rt-breakdown}
\end{figure}
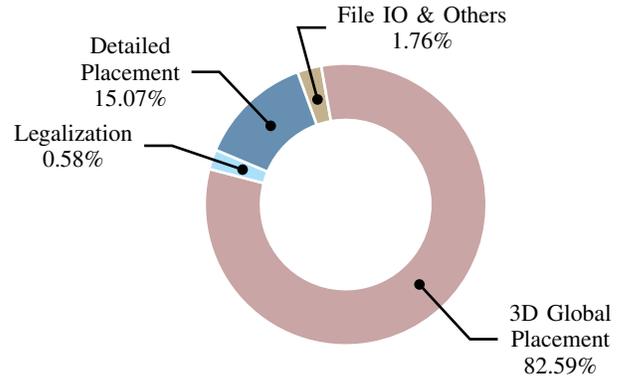
\Cref{fig:rt-breakdown} plots the overall runtime breakdown on the benchmark \texttt{case4h} for our 3D analytical placement framework. The GPU-accelerated 3D global placement takes 82.59\% of the total runtime, while the detailed placement takes 15.07\%.

Similar to~\cite{lin2019dreamplace}, the density and its gradients are computed with a GPU-accelerated implementation of 3D FFT in the 3D global placement. Given the ultra-fast density computation, we set the number of bins $N_z=\text{32}$ by default for all nontrivial cases in~\cite{hu20222022} so that the discrete grids can model the 3D electric field more precisely and thus produce better results. The proposed bistratal wirelength model is also implemented based on weighted-average~\cite{hsu2011tsv,hsu2013tsv} with GPU-acceleration techniques in~\cite{lin2019dreamplace}. The computation of wirelength and density with their gradients take up the main part of runtime in global placement.
It is worth mentioning that we can achieve 9.807$\times$ and 7.506$\times$ runtime speedup for the largest two designs \texttt{case4} and \texttt{case4h} over the first-place winner, demonstrating that our placement framework is scalable.

\section{Conclusions}
\label{sec:conclusion}
This paper proposes a new analytical 3D placement framework for face-to-face (F2F) bonded 3D ICs with heterogeneous technologies, incorporating a novel bistratal wirelength model.
The proposed framework leverages high-performance GPU-accelerated implementations of both the wirelength model and the electrostatic-based density model.
The experimental results on ICCAD 2022 Contest benchmarks demonstrate that our framework significantly surpass the first-place winner and the SOTA analytical 3D placer by 4.1\% and 2.1\% on wirelength, respectively, with much fewer vertical interconnections and conspicuous acceleration.
The 3D placement framework accomplishes partitioning and placement in a single run, proving that true 3D analytical placement can effectively handle partitioning with respect to wirelength optimization for F2F-bonded 3D ICs and thus inspire more explorations and studies on 3D analytical placement algorithms.

{
  \bibliographystyle{IEEEtran}
  \bibliography{ref/Top,ref/all}
}

\end{document}